\documentclass[10pt]{elsarticle}
\usepackage{graphicx}
\usepackage{caption}
\usepackage{subcaption}
\usepackage{amsmath}
\DeclareMathOperator{\sgn}{sgn}
\usepackage{amssymb}
\usepackage{amsfonts}
\usepackage{siunitx}
\usepackage[margin=2cm]{geometry}
\usepackage{url}
\usepackage{bm}
\usepackage{lineno,hyperref}
\modulolinenumbers[5]
\journal{Astroparticle Physics Journal}
\begin{document}
\begin{frontmatter}
\title{Complex Analysis of Askaryan Radiation: A Fully Analytic Treatment including the LPM effect and Cascade Form Factor}
\tnotetext[mytitlenote]{Open-source code associated with this work:  https://github.com/918particle/AskaryanModule}

\author{Jordan C. Hanson\corref{corr}}
\address{Center for Cosmological and AstroParticle Physics, Department of Physics, The Ohio State University, Columbus, OH 43210}
\cortext[corr]{Corresponding author}
\ead{hanson.369@osu.edu}

\author{Amy L. Connolly\corref{corr1}}
\address{Center for Cosmological and AstroParticle Physics, Department of Physics, The Ohio State University, Columbus, OH 43210}

\begin{abstract}
The Askaryan effect describes coherent electromagnetic radiation from high-energy cascades in dense media with a collective charge.  We present an analytic model of Askaryan radiation that accounts simultaneously for the three-dimensional form factor of the cascade, and quantum mechanical cascade elongation via the Landau-Pomeranchuk-Migdal effect.  These calculations, and the associated open-source code, allow the user to avoid computationally intensive Monte Carlo cascade simulations.  Searches for cosmogenic neutrinos in Askaryan-based detectors benefit from computational speed, because scans of Askaryan parameter-space are required to match neutrino signals.  The Askaryan field is derived from cascade equations verified with Geant4 simulations, and compared with prior numerical and semi-analytic calculations.  Finally, instructive cases of the model are transformed from the Fourier domain to the time-domain.  Next-generation \textit{in situ} detectors like ARA and ARIANNA can use analytic time-domain signal models to search for correlations with event candidates.
\end{abstract}


\end{frontmatter}


\section{Introduction}
\label{sec:intro}

The landmark observation of PeV neutrino interactions in Antarctic ice by the IceCube collaboration \cite{PhysRevLett.111.021103} has highlighted the urgency for progress in ultra-high energy cosmogenic neutrino (UHE-$\nu$) searches, at energies $10^{16}-10^{20}$ eV \cite{PhysRevD.66.063004} \cite{1475-7516-2010-10-013} \cite{LearnedManheim}.  Cosmogenic neutrinos represent a long awaited prize in both astrophysics and particle physics, because of the potential to explain the origin of UHE cosmic rays (UHECR), as well as the chance to study electroweak interactions at record-breaking energies.  

The GZK process is a $p\gamma$ interaction yielding UHE-$\nu$ from the UHECR flux through the cosmic microwave background (CMB) \cite{PhysRevLett.16.748} \cite{zatsepin1969pis}.  UHECR models lead to the conclusion that 100 km$^3$-volume detectors are required to measure cosmogenic UHE-$\nu$ flux \cite{PhysRevD.59.023002} \cite{PhysRevD.64.093010} \cite{PhysRevD.88.121301} \cite{Joshi21042014}.  UHE-$\nu$ also present the possibility revealing physics beyond the Standard Model, via measurements of UHE-$\nu$ deep-inelastic scattering cross-sections \cite{PhysRevD.83.113009} \cite{PhysRevD.86.103006}.  Models matching simultaneously the PeV-$\nu$ flux from IceCube, the diffuse GeV gamma-ray flux, and the UHECR flux, are now placing constraints on cosmogenic UHE-$\nu$ flux \cite{PhysRevD.88.121301} \cite{Ahlers2010106} \cite{PhysRevLett.116.071101}.

The next generation of UHE-$\nu$ detectors is based on the Askaryan effect, in which a UHE-$\nu$ interaction produces a cascade that radiates radio-frequency (RF) pulses from within a dielectric medium \cite{askaryan1962excess} \cite{PhysRevLett.86.2802} \cite{PhysRevLett.99.171101}.  These detectors use the special RF properties of Antarctic ice, in order to search for UHE-$\nu$ cascades efficiently \cite{2011ICRC} \cite{HansonMoore} \cite{Besson2008130}.  The Radio Ice Cernekov Experiment (RICE) conducted the pioneering search \cite{PhysRevD.85.062004}.  The Antarctic Impulsive Transient Antenna (ANITA) is a balloon-borne detector \cite{PhysRevD.82.022004} \cite{Gorham200910}.  The Askaryan Radio Array (ARA), and the Antarctic Ross Ice Shelf Antenna Neutrino Array (ARIANNA) are two \textit{in situ} detectors similar to RICE, but designed on a much larger scale \cite{Barwick201512} \cite{Barwick2015139} \cite{7283676} \cite{Allison201562} \cite{Allison2012457}.  The ExaVolt Antenna (EVA) is a proposed design to improve on the ANITA detection scheme \cite{Gorham2011242}.

The expected Askaryan RF pulse in the detectors must be understood in detail.  Zas, Halzen, and Stanev (ZHS) created a Monte Carlo (MC) simulation which yielded the Askaryan field by tracking the radiation from every cascade particle above $\approx 1$ MeV, using the Fraunhofer approximation \cite{PhysRevD.45.362}.  This technique is computationally intensive and is difficult to scale to UHE-$\nu$ energies.  Semi-analytic models \cite{PhysRevD.61.023001} \cite{PhysRevD.84.103003} \cite{PhysRevD.81.123009} by J. Alvarez-Muniz, A. Romero-Wolf, R.A. Vazquez, and E. Zas (AVZ, ARVZ) solve Maxwell's equations, treating the cascade charge excess as a source current.  These models require only the MC \textit{profile} (charge versus depth) of the cascade.  Semi-classical methods become fully analytic when the Greisen and Gaisser-Hillas treatments provide the profile \cite{Eug}.

J. Ralston and R. Buniy (RB) \cite{PhysRevD.65.016003} presented a complex analysis of Askaryan radiation.  This approach yields theoretical insight into observable properties of the field, while matching the ZHS MC.  The model includes an explanation of signal causality, and merges coherence zones (Near-Field, Fresnel, Fraunhofer) continuously.  A proper handling of non-Fraunhofer zones is vital for lowering energy thresholds in the \textit{in situ} Askaryan detectors, because the lowest-energy signals are above thermal backgrounds when the neutrino vertex is in coherence zones other than the Fraunhofer zone.  Lowering energy thresholds is critical for maximizing detected UHE-$\nu$, because the flux is expected to increase with decreasing energy.

We generalize the RB model by including two effects.  We derive and implement the three-dimensional cascade form factor from lateral MC charge diffusion, $\widetilde{F}(\omega,\theta)$, and we introduce smooth cascade-elongation from the Landau-Pomeranchuk-Migdal (LPM) effect.  Among the consequences of these extensions to RB are the constraint of the high-frequency ($\approx 1$ GHz) Fourier modes and the solid angle of the radiation.  Parameters in $\widetilde{F}(\omega,\theta)$ are derived from Geant4 MC ($>1$ PeV) using a sub-cascade technique, and the electromagnetic MC cascades match the Greisen treatment, laterally and longitudinally \cite{GREISEN426}.  Finally, the functional form of the Askaryan field is derived for special cases useful for Askaryan signal template-creation for UHE-$\nu$ searches with ARA/ARIANNA.

\subsection{Advantages of a Fully Analytic Treatment with the Form Factor and LPM Elongation}

The challenge of detecting UHE-$\nu$ with ARA/ARIANNA is one of distinguishing natural RF pulses from backgrounds.  Every RF antenna observes thermal radiation associated with the temperature of the observed medium.  ARA/ARIANNA background rates are a mixture of thermal radiation from ice, the sky, and the Milky Way \cite{Barwick201512} \cite{Allison2012457} \cite{PhysRevD.93.082003}.  Thermal noise is uncorrelated with digital \textit{signal templates} generated by convolving theoretical Askaryan pulses with detector response.  Signal templates have been used recently to detect UHECR with the ARIANNA Hexagonal Radio Array (HRA) \cite{Barwick201750}.  More significantly, vibrational signal templates were used to detect gravitational wave signals in the Advanced LIGO detector \cite{PhysRevLett.116.061102}.  In each case, a \textit{template library} was created, in which individual templates share a physical origin, but use different functional parameters that span parameter-space.

Creating template libraries with semi-analytic and fully analytic models is far more efficient than creating them with particle MC codes.  The Askaryan field is derived in semi-analytic models from cascade profiles computed via MC codes, accounting for stochastic effects in cascade profiles \cite{PhysRevD.61.023001}.  Fully analytic models are derived mathematically, where every model parameter has a physical origin and explanation.  Template libraries may then be generated by tuning continuously parameters, accounting for regions of parameter-space where the form factor $\widetilde{F}(\omega,\theta)$ and LPM elongation dominate.  Although the resulting template library would not account for stochastic processes, it would contain the effects of the form factor and LPM elongation, which control the sensitivity and design of Antarctic detectors.  Examples of parameter-space variables are: the range to the interaction $R$, the frequencies $\nu$, the length of the cascade $a$, the viewing angle $\theta$, the fraction of the cascade that is electromagnetic or hadronic, the total cascade energy $E_{\rm C}$, and the lateral width of the cascade $\sqrt{2\pi}\rho_{\rm 0}$.

\section{Units, Definitions, and Conventions}
\label{sec:unit}

All calculations in this work have been encapsulated into an open-source C++ class, available online \footnote{https://github.com/918particle/AskaryanModule}.  The primary function of this code is to predict the electric fields that Askaryan-based detectors would detect.  In all sections, this class will be called the \textit{associated code}, or simply \textit{the code}.  

\begin{figure}
\centering
\begin{subfigure}{0.4\textwidth}
\centering
\includegraphics[width=0.55\textwidth,angle=270,trim = 0cm 1cm 1cm 1cm,clip=true]{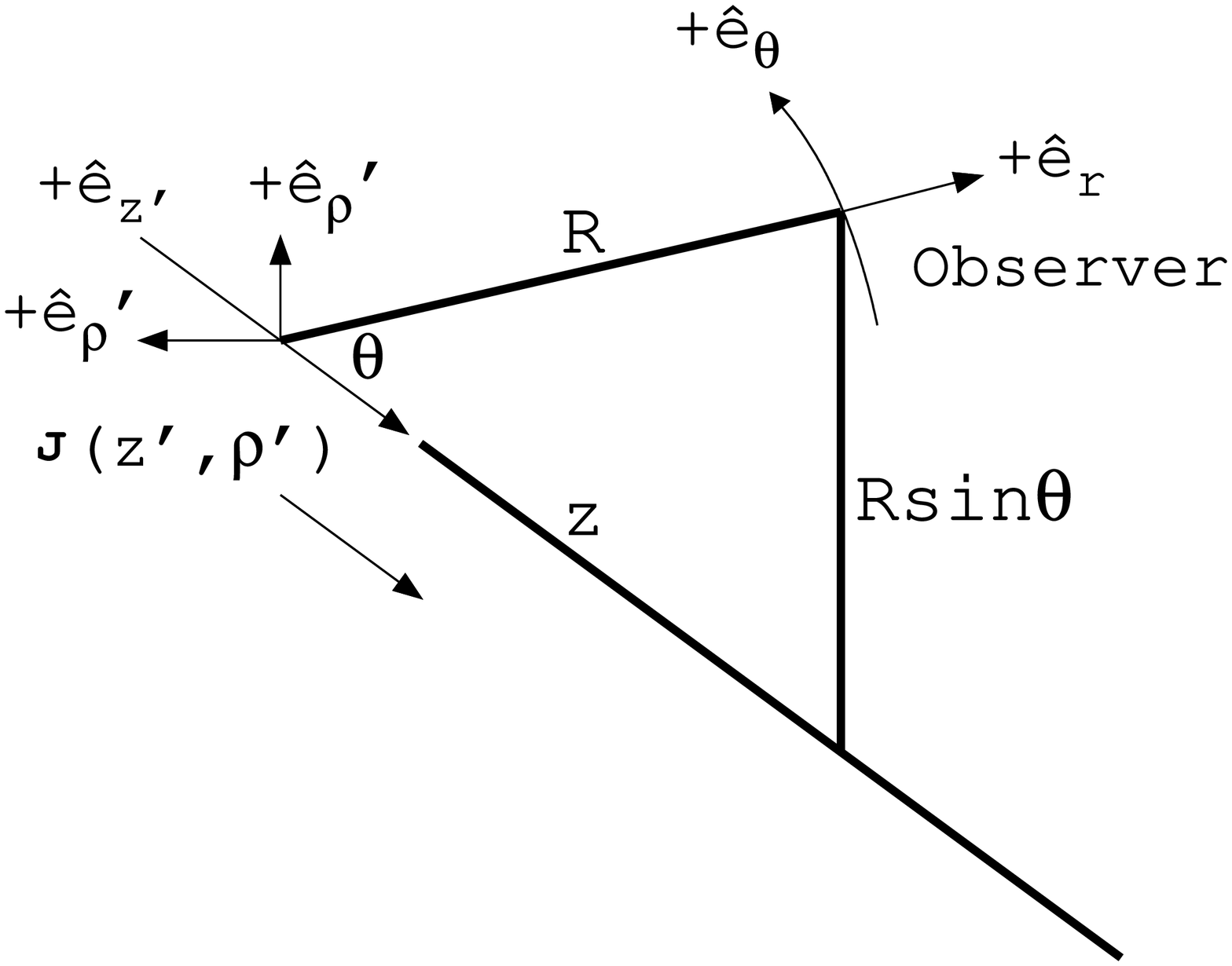}
\caption{}
\end{subfigure}%
\begin{subfigure}{0.4\textwidth}
\centering
\includegraphics[width=0.9\textwidth,trim = 0cm 7cm 0cm 7cm,clip=true]{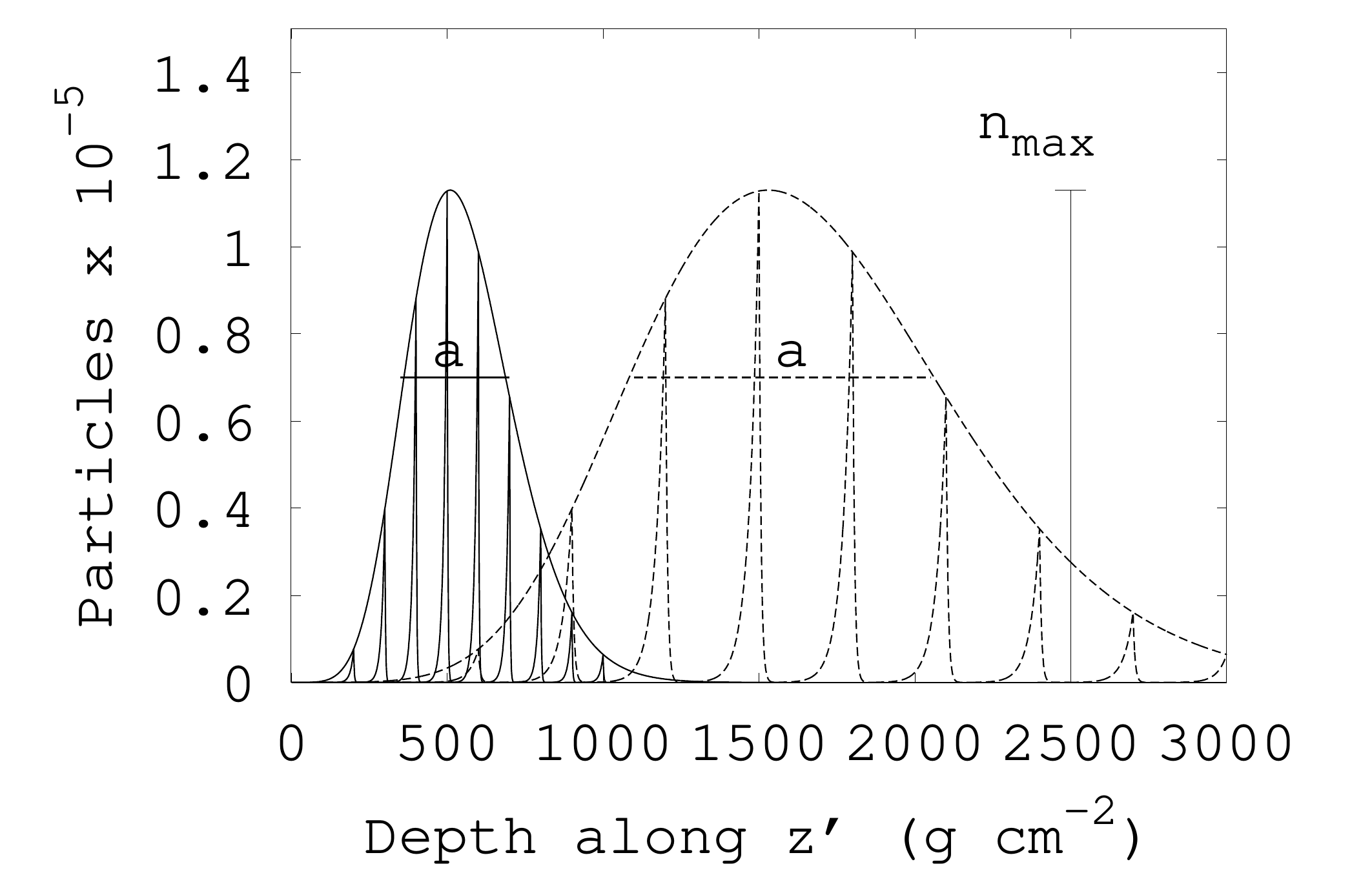}
\caption{}
\end{subfigure}
\caption{\label{fig:coord} (a) Coordinate systems.  Un-primed coordinates refer to the observer, and primed coordinates refer to the reference frame of the vector current.  (b) The curves represent the total charge content, with the instantaneous charge density (ICD) shown by the internal curves, versus depth after the first interaction in the medium.  The solid curves were made for shower energy $10^{14}$ eV and the dashed curves for a shower profile having been stretched by either the LPM effect, or simply by going to a higher energy.  The Greisen model was used for the form of the cascade profile.  The $n_{\rm max}$ parameter will refer to the \textit{negative} charge excess.}
\end{figure}

The coordinate systems are shown in Fig. \ref{fig:coord}a.  Observer coordinates are un-primed, and charge excess coordinates are primed.  The cascade current $\textbf{J}(t')$ is described in Sec. \ref{sec:rb0}, and will be called the instantaneous charge distribution (ICD) in subsequent sections.  The vectors $\rho$ and $\rho'$ refer to the lateral distance from the cascade axis, and $z$ and $z'$ refer to the cascade axis.  The origin for both systems is the location of the cascade maximum, with $z' = z = 0$ and $\rho = \rho' = 0$.  The viewing angle is $\theta$.  Bold variables, and variables with a circumflex, $\hat{e}_i$, refer to vectors.  The observer distance is $R = |\textbf{x}-\textbf{x}'|$, $\omega$ refers to the angular frequency, and $k=(2\pi)/n\lambda$ and $\mathbf{q} = n(\omega R,\omega \rho)/(cR)$ refer to one- and three-dimensional wavevectors \textit{in the dielectric}, with a refractive index $n$.  Although the code takes $n=1.78$ as a default value, it may be altered to apply to other uniform media.  The Cherenkov angle is defined by the index of refraction: $\cos\theta_{\rm C} = 1/n$.  The index of refraction at RF frequencies for bulk ice in Antarctica is $1.783\pm 0.003$ \cite{Bogo}.

The archetypal cascade profile is shown in Fig. \ref{fig:coord}b, in which the total number of cascade particles is shown, versus depth along the $z'$ axis.  The depth is $P z'$, where $P$ is the bulk density of ice (0.917 g/cm$^3$).  The parameter $n_{\rm max}$ will refer to the net number of \textit{negatively charged} particles (typically $\approx$ 10\% to 20\% of the total - see Sec. \ref{sec:form0} and the Appendix Sec. \ref{sec:form1} for details).  The parameter $a$ is the Gaussian width of the cascade profile near cascade maximum.  The cascade is initiated by an electroweak neutrino interaction, where the neutrino energy is $E_{\rm \nu}$, and the total energy of the cascade is $E_{\rm C}$.

The units of the electromagnetic field in the Fourier domain are $\SI{}{\V/\m/\Hz}$, often converted in the literature to $\SI{}{\V/\m/\MHz}$.  To make the distance-dependence explicit, both sides of field equations are multiplied by $R$, making the units $\SI{}{\V/\Hz}$.  In Sec. \ref{sec:time}, the field normalization constant is $E_{\rm 0}$, and it contains the energy-dependence.  $E_{\rm 0}$ may be linearly scaled with energy, provided the parameters $a$ and $n_{\rm max}$ are derived consistently from the Greisen model.  Equations in Sec. \ref{sec:time} are proportional to $\omega E_{\rm 0}$, so the units of $E_{\rm 0}$ are $\SI{}{\V/\Hz^2}$.

In the original RB work, the following convention for the Fourier transform $\tilde{G}(\omega)$ of a function $g(t)$ was used:

\begin{align}
\label{eq:FT}
\tilde{G}(\omega) &= \int_{-\infty}^{\infty} e^{i\omega t} g(t) dt \\
\label{eq:FT2}
g(t) &= \frac{1}{2\pi} \int_{-\infty}^{\infty} e^{-i\omega t} \widetilde{G}(\omega)
\end{align}

The sign convention shown in the exponent is used in RB, though the opposite sign convention is more common in mathematical physics.  The definitions in Eqs. (\ref{eq:FT}, \ref{eq:FT2}) have been kept for consistency with Sec. V of \cite{PhysRevD.65.016003}.  The sign convention may be toggled in the code, but the output does not depend on this choice, because the appropriate transformation in time is applied.  All frequencies are shown in MHz in spectral plots, for comparison to the literature.  The symbol $\sim$ above a function denotes a Fourier-domain quantity.  In Secs. \ref{sec:rb0} and \ref{sec:form}, the three-dimensional Fourier transform is used to describe the form factor in three-dimensional frequency-space, with the normal sign convention:

\begin{equation}
\label{FF}
\widetilde{F}(\textbf{q}) = \int d^3 x' f(\textbf{x}') e^{-i \textbf{q}\cdot \textbf{x}'}
\end{equation}

\section{Combination of the RB model, the LPM Effect, and the Cascade Form Factor}

In Sec. \ref{sec:assume}, the RB model \cite{PhysRevD.45.362} is presented for clarity, beginning with all assumptions made.  Next, the model is enhanced by treating two new effects: \textit{the LPM effect} (Sec. \ref{sec:lpm}) and \textit{the cascade form factor}, $\widetilde{F}$ (Sec. \ref{sec:form}).

\subsection{General Assumptions, and the Basic RB Model}
\label{sec:assume}

The conventions are taken from Sec. \ref{sec:unit} to build the RB model in the code \cite{PhysRevD.65.016003}.  RB derive the Askaryan fields from Maxwell's equations in the Lorentz gauge for a dielectric medium, by expanding with a special scalar parameter $\eta$.  The source current, $\mathbf{J}(t',\textbf{x}')$, is described by an ICD $f(z'-vt,\textbf{$\rho$}')$, excess charge profile $n(z')$, and cascade speed $\mathbf{v}$:

\begin{align}
\label{eq:new0}
\mathbf{J}(t',\textbf{x}') &= \textbf{v}n(z')f(z'-vt,\textbf{$\rho$}') \\
\label{eq:new1}
\eta &= \left(\frac{a}{\Delta z_{\rm coh}}\right)^2 = \frac{k}{R} (a\sin\theta)^2
\end{align}

The function $f(z'-vt,\textbf{$\rho$}')$ is the 3D ICD.  Equation \ref{eq:new1} is the squared ratio of $a$, and $\Delta z_{\rm coh}$, the longitudinal range where the Askaryan radiation must be coherent.  To understand $\Delta z_{\rm coh}$, consider Feyman's formula \cite{PhysRevD.45.362}.  Radiation from an accelerating point charge is proportional to the angular acceleration $\ddot{\theta}$ relative to the observer.  The coherence regime is defined by $|z'| \lesssim \Delta z_{\rm coh}$, where $\ddot{\theta}$ and $R(z')$ are constant in time to first order, and $\ddot{\theta}$ is maximized.  RB show that if $\Delta R(z') < \lambda$ in this limit, $\Delta z_{\rm coh} < (R/(k\sin^2\theta))^{1/2}$.

The dominant Askaryan radiation occurs when $|z'| \lesssim \Delta z_{\rm coh}$.  In the Fresnel and Fraunhofer (far-field) regimes, $\eta < 1$.  The calculations, however, are valid for any $\eta$, rather than only the in the far-field ($\eta \rightarrow 0$, and $kR \gg 1$, $R \gg a$).  If $a \ll \Delta z_{\rm coh}$, then the fields have spherical symmetry, and the limit $kR \gg 1$ corresponds to the far-field.  Conversely, if $a \gg \Delta z_{\rm coh}$, then $\eta \rightarrow \infty$, the fields have cylindrical symmetry.

The longitudinal cascade width $a$, and therefore $\eta$, is derived in the code from either the Greisen (electromagnetic) or Gaisser-Hillas (hadronic) cascade profile functions (Secs. \ref{sec:form}-\ref{sec:form0}), and can be elongated by the LPM effect.  The associated code requires the type of cascade profile (hadronic or electromagnetic) as an input, and whether the LPM effect must be applied.  If the user specifies a purely hadronic cascade, no further action is taken if the LPM effect is activated \cite{PhysRevD.61.023001}.

\subsection{The RB Field Equations}
\label{sec:rb0}

RB insert the vector current $\textbf{J}(t',\textbf{x}')$ into Maxwell's equations, and solve for the vector potential:

\begin{equation}
\label{eq:rb1}
c \textbf{A}(\textbf{x}') = \int d^3 x' \frac{e^{ik|\textbf{x}-\textbf{x}'|}}{|\textbf{x}-\textbf{x}'|} \int dt' e^{i\omega t'} \textbf{J}(t',\textbf{x}')
\end{equation}

RB then define $R(z') = \sqrt{(z-z')^2+\rho^2})$, and expand around $\rho' = 0$:

\begin{equation}
\label{eq:rb2}
|\mathbf{x}-\mathbf{x}'| \approx R(z') - \frac{\rho \cdot \rho'}{R(z')} + \frac{\rho'^2}{2 R(z')}
\end{equation}

In Eq. \ref{eq:rb2}, the third term on the right-hand side is dropped, because $\rho' \ll R(z')$.  The vector potential in Eq. \ref{eq:rb1} is then factored into the form factor $\widetilde{F}(\omega)$ and a vector potential.  $\widetilde{F}(\omega)$ is the three-dimensional Fourier transform of the ICD, and the vector potential is governed by $n(z')$.  \textit{Thus, $\widetilde{F}(\omega)$ describes the charge distribution, and the vector potential describes the charge evolution}.  Equations \ref{eq:rb3}-\ref{eq:rb3a} summarize the result.  In Eqs. \ref{eq:rb3}-\ref{eq:rb3a}, $\mathbf{\widetilde{A}}^{FF}$ is the vector potential, named the \textit{Fresnel-Fraunhofer} (FF) potential in RB.

\begin{align}
\label{eq:rb3}
\mathbf{\widetilde{A}}(\omega,\theta) &=\mathbf{\widetilde{A}}^{FF} (\omega,\theta) \int d^3x' e^{-i\textbf{q}\cdot \textbf{x}'} f(\textbf{x}') \\
\label{eq:rb3a}
\mathbf{\widetilde{A}}(\omega,\theta) &= \widetilde{F}(\omega,\theta) \mathbf{\widetilde{A}}^{FF} (\omega,\theta)
\end{align}

Equations \ref{eq:money}-\ref{eq:money4} express the general RB result for the electric field $\textbf{E} = -\partial\textbf{A}/\partial t$, in terms of the frequency $\nu$, viewing angle $\theta$, and $\eta$:

\begin{align}
\label{eq:money}
\frac{R\mathbf{\widetilde{E}}(\nu,\theta,\eta)}{\left[\frac{\SI{}{\V}}{\SI{}{\MHz}}\right]} &= 2.52 \times 10^{-7} \frac{a}{\left[\SI{}{\m}\right]} \frac{n_{\rm max}}{\left[1000\right]} \frac{\nu}{\left[\SI{}{\GHz}\right]} \widetilde{F}(\textbf{q}) \psi \bm{\mathcal{E}} \\
\label{eq:money2}
\psi &= -i e^{ikR} \sin\theta \\
\label{eq:money3}
\bm{\mathcal{E}} &= \mathcal{W}(\eta,\theta)\left(\frac{\cos\theta_{\rm C} - \cos\theta}{\sin\theta}\right)\hat{e}_{\rm r} +  \mathcal{W}(\eta,\theta)\left(1-i\eta \frac{\cos\theta_{\rm C}}{\sin^2\theta} \frac{\cos\theta - \cos\theta_{\rm C}}{1-i\eta}\right)\hat{e}_{\rm \theta} \\
\label{eq:money4}
\mathcal{W}(\eta,\theta) &= \left(1-i\eta\left(1-3i\eta\frac{\cos\theta}{\sin^2\theta}\frac{\cos\theta - \cos\theta_{\rm C}}{1-i\eta}\right)\right)^{-1/2} \exp\left(-\frac{1}{2}(ka)^2 \frac{(\cos\theta-\cos\theta_{\rm C})^2}{1-i\eta}\right)
\end{align}

Equation \ref{eq:money} is the total field, with an overall phase factor defined in Eq. \ref{eq:money2}.  Equation \ref{eq:money3} contains the vector structure, and Eq. \ref{eq:money4} governs the phase and angular structure.

\subsection{The Landau-Pomeranchuk-Migdal (LPM) Effect}
\label{sec:lpm}

The Landau-Pomeranchuk-Migdal (LPM) effect is a suppression of the pair-creation and bremsstrahlung cross-sections at cascade energies above a material-dependent constant known as the LPM energy or $E_{\rm LPM}$ \cite{RevModPhys.71.1501}:

\begin{equation}
\label{eq:lpm2}
E_{\rm LPM} = \frac{(m c^2)^2 \alpha X_{\rm 0}}{4\pi c \hbar} = 7.7 ~ \SI{}{\TeV}/\SI{}{cm} \cdot X_{\rm 0}
\end{equation}

In Eq. \ref{eq:lpm2}, $m$ is the electron mass, $\alpha$ is the fine-structure constant, and $X_{\rm 0}$ is the radiation distance ($E_{\rm LPM} = 0.303$ PeV for ice).  If $E_{\rm C} > E_{\rm LPM}$, the quantum mechanical \textit{formation length} of these interactions is longer than atomic separations, leading to quantum interference and suppressing cross-sections.  The result is a  \textit{longitudinal shower elongation}, from the non-interacting, higher-energy particles.  The LPM effect is reviewed by S.R. Klein in Ref. \cite{RevModPhys.71.1501}.

The LPM changes how the RB model must be applied.  Under normal circumstances, the quantity $n_{\rm max} a$ approximates the area under the cascade profile for particles with energy greater than some critical energy: $n_{\rm max} a \propto E_{\rm C}$.  It is shown in Ref. \cite{101103physrevd82074017} that in the LPM-regime, $a \propto \sqrt{E_{\rm C}/E_{\rm crit}}$.  While it may appear that changing the energy scaling of $a$ violates energy conservation through the field normalization $n_{\rm max} a$, this is not necessarily the case.

First, the amount of \textit{radiated} energy does not have to track the \textit{total} energy, if the cascade shape is modified.  The Frank-Tamm radiation distribution for a charge moving through a track of length $L$ states: $d^2 P/d\Omega d\omega \propto L^2$ \cite{PhysRevD.65.016003}.  The LPM effect would increase $L$, increasing the radiated energy.  The increase is contingent on $L$ remaining comparable to $\Delta z_{\rm coh}$, avoiding spectral cut-off.  It is shown in Sec. \ref{sec:rb1} (graphically) and Sec. \ref{sec:time} (mathematically) that enhancements due to cascade elongation are attenuated by a combination of decoherence, $\widetilde{F}$, and narrowed Cherenkov cone-width.

Second, it should not be assumed naively that $n_{\rm max}$ must be decreased in direct proportion to the increase in $a$, under LPM elongation.  Absent the LPM effect, the inverse relationship between $a$ and $n_{\rm max}$ merely reflects that $E_{\rm C}$ is a constant, with $\approx 10^{-7}$ of the energy in TeV passing to Askaryan radiation in V/m/MHz.  Figures 12-13 of Ref. \cite{Cillis_Influence_1999} demonstrate LPM elongation, but $n_{\rm max}$ is not reduced in direct proportion to $a$.  Figure 12 of Ref. \cite{Cillis_Influence_1999} shows that $X_{\rm max}$ grows and fluctuates at energies above $E_{\rm LPM}$.  However, Fig. 13 of Ref. \cite{Cillis_Influence_1999} shows that the energy-dependence of $n_{\rm max}$ stays approximately the same, and $n_{\rm max}$ fluctuates mildly.  Further, Fig. 3 of \cite{PhysRevD.61.023001} suggests that charged-current induced cascades, with mixed hadronic and electromagnetic components, have no reduction in $n_{\rm max}$ relative to neutral-current induced cascades, for which the LPM effect is not relevant \cite{AlvarezMuniz1998396}.

Physically, all particle energies must decrease below $E_{\rm LPM}$ eventually.  This takes place before the cascade maximum, so to first order $n_{\rm max}$ is still governed by non-LPM physics, even if $E_{\rm C} > E_{\rm LPM}$.  Conversely, ZHS and others have shown that the radiated field strength below 500 MHz is proportional to the weighted, projected track length of charged particles \cite{PhysRevD.45.362}.  The LPM effect cannot enhance dramatically the weighted projected track length, because the number of LPM-elongated tracks is a small fraction of the total number.  The associated code contains a flag (\textit{strictLowFreqLimit}) which directs the model to retain this limit by inversely scaling $n_{\rm max}$ in direct proportion to the increase in $a$.  Requiring this limit is equivalent to stating that sub-showers with $E<E_{\rm LPM}$ occur both \textit{before} and \textit{after} cascade maximum, \textit{smearing} $n(z')$.  From the arguments and references above, this is the most conservative approach.

Finally, S.R. Klein notes in Ref. \cite{RevModPhys.71.1501} that the LPM effect may influence $\widetilde{F}(\omega,\theta)$.  The lateral ICD is influenced by multiple scattering (MSC) effects that lead to a mean scattering angle $\langle \theta_{\rm MSC} \rangle$ for cascade particles incident on atoms \cite{RevModPhys.71.1501} \cite{GREISEN426}.  In Ref. \cite{RevModPhys.71.1501}, $\langle \theta_{\rm MSC} \rangle$ takes the following form for a particle of energy $E$:

\begin{equation}
\langle \theta_{\rm MSC} \rangle = \frac{E_{\rm s}}{E} \sqrt{\frac{d}{X_{\rm 0}}}
\label{eq:mscAngle}
\end{equation}

$X_{\rm 0}$ is the radiation distance, $d$ is the distance over which the scattering occurs, and $E_{\rm s}$ is the Moli\`{e}re scattering energy ($E_{\rm s} = mc^2 \sqrt{4\pi/\alpha} \approx 21.2$ MeV).  The LPM effect increases $d$ relative to $X_0$, increasing $\langle \theta_{MSC} \rangle$.  Although $\langle \theta_{\rm MSC} \rangle $ is enhanced by the LPM effect, it is also inversely proportional to energy. This implies that the pile-up of particle energies at $E_{\rm crit} \approx E_{\rm s}$ actually governs the ICD width.

Finally, a remark about ``shower fluctuations" is prudent.  While the LPM effect may cause the cascade maximum location to fluctuate, the \textit{location} of cascade maximum is irrelevant, because Askaryan radiation is independent of the origin of the coordinate system.  The origin in the associated code is the location of cascade maximum, \textit{wherever it occurs}.  Stochastic features of the LPM effect, such as the multi-peaked cascade profile, are not treated in this work.  However, it is shown in Ref. \cite{Hu_Near_2012} that the sub-peaks only alter the waveform in certain circumstances, such as being off-cone by $\approx 10^{\circ}$ in the Fraunhofer regime.  The waveforms in Ref. \cite{Hu_Near_2012} all contain the basic bi-polar structure produced by the associated code, despite the multi-peaked cascade profile.

\begin{figure}
\begin{center}
\includegraphics[width=0.55\textwidth,trim=0cm 6.5cm 0cm 6.5cm,clip=true]{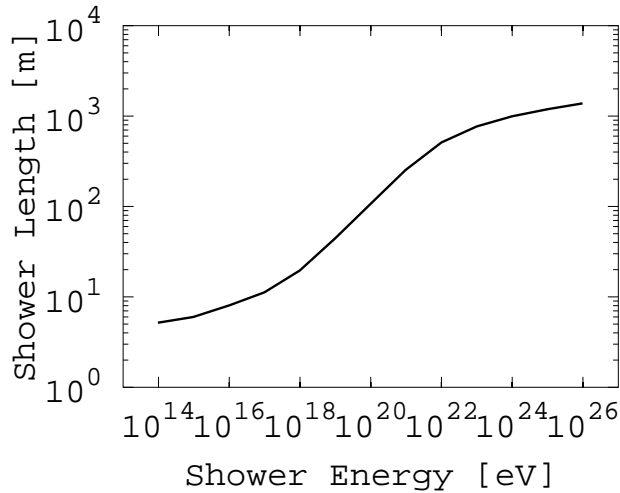}
\end{center}
\caption{\label{fig:length} The length of the neutrino-induced cascade, including the LPM effect, from ref. \cite{101103physrevd82074017}.}
\end{figure}

In summary, the main effect LPM physics has on Askaryan radiation is the angular and frequency filtering via the cascade elongation.  The effect is quantified in the code by drawing an elongated shower width $a$ from calculations by Klein and Gerhardt shown in Fig. \ref{fig:length} \cite{101103physrevd82074017}.  In scenarios where LPM is unimportant, such as initial hadronic-dominated processes \cite{AlvarezMuniz1998396} and cascades with $E_{\rm C} < E_{\rm LPM}$, the associated code draws the width and height of the negative charge excess profile from the usual Greisen and Gaisser-Hillas formulations \cite{GREISEN426} \cite{Gaisser}.

\subsection{The Cascade Form Factor, $\widetilde{F}$}
\label{sec:form}

The factorization of the longitudinal charge excess \textit{evolution}, $\mathbf{\widetilde{A}}^{FF} (\omega,\theta)$, and the instantaneous properties of the lateral charge \textit{distribution}, $\widetilde{F}(\omega)$, leads to the interpretation of $\widetilde{F}(\omega)$ as a filter.  Filters are fully described by \textit{pole-zero} diagrams, which display the Laplace transform of the filter transfer function.  Impulsive E-fields with no DC-component automatically approach 0 as $|\omega| \rightarrow 0$, meaning $\widetilde{F}$ should not require any zeros.  Thus, $\widetilde{F}(\omega)$ should be completely defined by poles in the complex $\omega$-plane.

D. Garcia-Fernandez \textit{et. al.} (Ref. \cite{Garcia_2012}) show that the integrals over individual tracks in the ZHS algorithm can be generalized to all coherence regimes, when integrated numerically.  The authors of Ref. \cite{Garcia_2012} then insert the form factor chosen in Ref. \cite{YuHu} into the RB framework to demonstrate agreement (Fig. 7 of Ref. \cite{Garcia_2012}).  This agreement occurs at frequencies below 120 MHz, below the bandwidths of ARA/ARIANNA.  The lateral $\rho'$-dependence of the chosen form factor is Gaussian ($f \propto \exp(-\rho'^2)$), yielding Gaussian behavior in the Fourier domain ($|\widetilde{F}| \propto \exp(-\omega^2)$).  Geant4 simulations and the Greisen model show that $f(\mathbf{x}')$ is not gaussian (Sec. \ref{sec:form0} and Appendix Sec. \ref{sec:form1}). Additionally, the ultra-violet divergence in Ref. \cite{Garcia_2012} is attributed to the assumption that the dielectric constant $\epsilon(\omega)$ is not absorptive, however, the effect of $\widetilde{F}(\omega)$ is apparent in that model near 1 GHz.  This is because the filtering effect of $\widetilde{F}(\omega,\theta)$ occurs when the Askaryan radiation of all finite tracks in a cascade are summed.

The ZHS form factor, $\widetilde{F}_{\rm ZHS}(\omega)$, is constructed of poles which imply that the lateral ICD $f \propto \exp(-\rho')$.  The key result for $\widetilde{F}(\omega)$ is given in Eq. \ref{eq:newForm}, in terms of the viewing angle $\theta$, the angular frequency $\omega$, and $(\sqrt{2\pi}\rho_0)^{-1}$, the distance from the cascade axis ($\rho' = 0$) at which the negative charge excess has decreased by $1/e$. 

\begin{equation}
\label{eq:newForm}
\widetilde{F}(\omega,\theta) = \frac{1}{\left(1+\left( \frac{(\omega/c) \sin\theta}{\sqrt{2\pi}\rho_{\rm 0}} \right)^2\right)^{3/2}}
\end{equation}

Intuitively, the squared ratio in the denominator of Eq. \ref{eq:newForm} compares the lateral projection of the wavevector with the physical extent of the charge excess.  If the charge excess is laterally large, compared to the wavelength, then Eq. \ref{eq:newForm} begins to act as a low-pass filter.  \textit{By substituting Eq. \ref{eq:newForm} into Eq. \ref{eq:money}, the RB model is completed, properly accounting for LPM elongation and the ICD.}  The form factor $\widetilde{F}(\omega,\theta)$ is derived from fits to Geant4 cascades, for $E_{\rm C} = 1-100$ PeV. Once the quality of these fits is established, one may proceed with the theoretical cascade models, with no need of further numerical simulation.

\subsubsection{The Greisen Model}
\label{sec:form0}

K. Greisen provided a comprehensive review of the longitudinal charge evolution, and lateral distribution, within electromagnetic cascades \cite{GREISEN426}, by combining earlier work by Moli\`{e}re, Nishimura and Kamata, Landau, and others.  The integrated charge at depth $z_0$ in the medium is

\begin{equation}
\label{eq:formG1}
n_{\rm tot}(z_{\rm 0}) = \frac{0.31}{\sqrt{\ln(E_{\rm C}/E_{\rm crit})}} \exp \left\lbrace z_{\rm 0}\left(1-\frac{3}{2} \ln(s) \right) \right\rbrace
\end{equation}

In Eq. \ref{eq:formG2}, $z_{\rm 0}$ is in units of radiation length (36.08 g cm$^{-2}$), and the cascade energy $E_{\rm C}$ in units of GeV ($E_{\rm crit} = 73$ MeV) \cite{PhysRevD.45.362,Eug}.  The \textit{shower age} is $s = 3z_{\rm 0}/(z_{\rm 0}+2\ln(E_{\rm C}/E_{\rm crit}))$.  Equation \ref{eq:formG1} contains the cascade maximum $z_{\rm max} = \ln(E_{\rm C}/E_{\rm crit})$.  Additionally, $\langle \ln n_{\rm tot} \rangle \approx 0.7(s-1-3\ln s)$.  When $s=1$ at $z_{\rm max}$, $\langle \ln n_{\rm tot} \rangle = 0$, implying negligible $n_{\rm max}$ fluctuation.  Approximating $n_{\rm tot}(z_{\rm 0})$ by a Gaussian, $a \propto \sqrt{z_{\rm max}}$.  For $n_{\rm max} = n_{\rm tot}(z_{\rm max})$, $n_{\rm max} a \propto E_{\rm C}/E_{\rm crit}$.  The factor $n_{\rm max} a$ approximates the area under the peak of Eq. \ref{eq:formG1}.

The lateral ICD arises from multi-scattering effects (MSC processes in Geant4), with average scattering angle $\langle \theta_{\rm MSC} \rangle$ (Eq. \ref{eq:mscAngle}), and a Moli\`{e}re radius defined by $\rho_{\rm 1} = E_{\rm MSC}/E_{\rm c}$, in radiation lengths.  Large errors arise for angles $\theta \gg \langle \theta_{\rm MSC} \rangle$, or ($\rho' > \rho_{\rm 1}$), but only 1-10\% of particles have $\rho' > \rho_{\rm 1}$.  The $\langle \theta_{\rm MSC} \rangle$ is expressed in Eq. \ref{eq:formG2}, for particles of energy $\epsilon$ in electromagnetic cascades \cite{RevModPhys.71.1501}:

\begin{equation}
\label{eq:formG2}
\langle \theta_{\rm MSC} \rangle = \frac{E_{\rm MSC}}{\epsilon} \sqrt{z_{\rm 0}} = \frac{m_e c^2 \sqrt{4\pi z_{\rm 0}/\alpha}}{\epsilon} \approx \frac{21.2 \left[\SI{}{\MeV}\right]}{\epsilon} \sqrt{z_{\rm 0}}
\end{equation}

Equation \ref{eq:formG2} implies the lateral ICD should be widest near $n_{\rm max}$, where $\epsilon$ is minimized.  The widening lateral ICD causes most particles to lag behind the cascade front by $\mathcal{O}(1-10)$ cm at $n_{\rm max}$ (see Eq. 13 of \cite{RevModPhys.71.1501}). Nishimura and Kamata refine the approximation, providing the lateral charged particle density, $D$.  The result is known as the NKG-function, shown in Eq. \ref{eq:formG3}.

\begin{equation}
\label{eq:formG3}
D = \frac{n_{\rm tot}}{2\pi \rho_{\rm 1}^2}\frac{\Gamma(4.5-s)}{\Gamma(s)\Gamma(4.5-2s)} \left( \frac{\rho'}{\rho_{\rm 1}} \right)^{s-2} \left( 1+\frac{\rho'}{\rho_{\rm 1}} \right)^{s-4.5}
\end{equation}

Agreement between this work and Eqs. \ref{eq:formG1}-\ref{eq:formG3} is shown Sec. \ref{sec:form1} in the Appendix.  The ICD is assigned a three-dimensional function $f(\mathbf{x}')$, and in Sec. \ref{sec:form2}, $\widetilde{F}(\omega)$ is derived analytically from $f(\mathbf{x}')$.  Other efforts to model the ICD and the resulting Askaryan radiation can be found in Refs. \cite{Razz} \cite{Alv2003} \cite{McKayHussain}.

\subsubsection{Analytic Formula for $\widetilde{F}(\omega,\theta)$}
\label{sec:form2}

The definition of $\widetilde{F}(\omega,\theta)$ is

\begin{equation}
\label{eq:form1}
\widetilde{F}(\textbf{q}) = \int d^3x' e^{-i\textbf{q}\cdot \textbf{x}'} f(\textbf{x}')
\end{equation}

The ICD $f(\mathbf{x}')$ is given by a general parameterization of the Greisen model:

\begin{equation}
\label{eq:form2}
f(\textbf{x}') = \rho_{\rm 0}^2 \delta(z') \exp(-\sqrt{2\pi}\rho_{\rm 0}\rho')
\end{equation}

The choice of Eq. \ref{eq:form2} is motivated by Sec. \ref{sec:form0} (see Appendix Sec. \ref{sec:form1} for more detail).  Recall that the ICD is meant to describe the number density of the negative charge excess, not the total charged particle number density.  Geant4 MC calculations and the Greisen model predict the exponential form, and the ZHS parameterization suggests it in the Fourier domain.  C.-Y. Hu \textit{et al} chose a double-Gaussian form \cite{YuHu}, which is not accurate near $\rho'=0$, but highlights the relationship between theoretical parameters and numerical fields.  Note that the units of the $1/e$ width $\sqrt{2\pi}\rho_0$ and $\delta(z')$ are inverse length, giving $f(\textbf{x}')$ units of number density.

Fitted results for the parameter $\sqrt{2\pi}\rho_{\rm 0}$ for times between 20-30 ns ($s\approx1$) indicate that $\sqrt{2\pi}\rho_{\rm 0} \sim$ is constant with respect to cascade depth.  The solution to Eq. \ref{eq:form1} is as follows: the trivial $z'$ integration is performed, setting $z' = 0$ without loss of generality.  Next, two convenient variables are defined, and shown in Eqs. \ref{eq:form3} and \ref{eq:form4}.

\begin{align}
\label{eq:form3}
\gamma &= \frac{\omega}{c}\sin\theta \\
\label{eq:form4}
\sigma &= \frac{\gamma}{\sqrt{2\pi}\rho_{\rm 0}}
\end{align}

The variable $\gamma$ is the lateral projection of the wavevector $\omega/c$, and $\sigma$ is the product of $\gamma$ and the lateral charge extent.  The variable $\sigma$ compares the laterally-projected wavelength to the lateral extent of cascade excess charge.  In Sec. \ref{sec:time}, $\sigma = \omega/\omega_{\rm CF}$, so that $\omega_{\rm CF}$ is the limiting frequency.  Substituting Eqs. \ref{eq:form2}, \ref{eq:form3} and \ref{eq:form4} into Eq. \ref{eq:form1}, $\widetilde{F}(\omega,\theta)$ becomes

\begin{equation}
\label{eq:form5}
\widetilde{F}(\omega,\theta) = \rho_{\rm 0}^2 \int_{\rm 0}^\infty d\rho' \rho' \exp\left\lbrace -\frac{\gamma}{\sigma}\rho' \right\rbrace \int_{-\pi}^\pi d\phi' \exp\left\lbrace -i\rho'\gamma \cos(\phi') \right\rbrace
\end{equation}

From the cylindrical symmetry of $f(\mathbf{x}')$, the $\phi'$ coordinate may be rotated.  With $\phi' \rightarrow \phi'-\pi/2$, the $\phi'$-integral becomes a 0th-order Bessel function.  A similar result in Ref. \cite{Razz} contains the Bessel function in Eq. \ref{eq:form5}.  In Ref. \cite{Razz}, however, the lateral ICD is not evaluated analytically, but through numeric integrals.  After making the substitution $u' = \gamma\rho'$, the remaining $\rho'$-integral may be found in standard tables.

\begin{align}
\label{eq:form6}
\widetilde{F}(\omega,\theta) &=\sigma^{-2} \int_0^\infty du' u' \exp\left\lbrace -\frac{u'}{\sigma} \right\rbrace J_{\rm 0}(u') \\
\label{eq:form7}
\widetilde{F}(\omega,\theta) &= \frac{1}{(1+\sigma^2)^{3/2}}
\end{align}

The result for $\widetilde{F}(\omega,\theta)$ is shown in Eq. \ref{eq:form7}.  The Askaryan spectrum is attenuated like $\omega^{-3}$ for $\sigma \gg 1$, for wavelengths much smaller than the lateral ICD.  For $\sigma \lesssim 1$, $\widetilde{F}(\omega,\theta) \approx (1+(3/2) \sigma^2)^{-1}$.  Given the the location of complex poles for $\sigma \ll 1$, one might suspect problems with causality.  It is important to note, upon transforming the model to the time domain, including the LPM effect and the effect of $\widetilde{F}(\omega,\theta)$, that the fields do not violate the causality criterion stated by RB \cite{PhysRevD.65.016003}.

Equation 16 in Ref. \cite{PhysRevD.84.103003} contains the Askaryan vector potential versus retarded time ($t_{\rm r})$, matched to MC at $\theta = \theta_C$, with six numerical parameters, not counting the overall normalization.  This equation is restated as Eq. \ref{eq:eq5a}, and the $x_{\rm i}$ have unique values for $t_{\rm r}>0$ and $t_{\rm r}<0$.

\begin{equation}
\frac{R \textbf{A}(t_{\rm r},\theta_{\rm C})}{\left[\SI{}{\volt}\cdot\SI{}{\second}\right]} = - E_{\rm 0}' \sin(\theta_{\rm C}) \hat{e_{\rm \theta}} \left(\exp(-2|t_{\rm r}|/x_{\rm 0})+(1+x_{\rm 1}|t_{\rm r}|)^{-x_{\rm 2}} \right)
\label{eq:eq5a}
\end{equation}

Equation \ref{eq:form7} fully describes the cascade shape, is analytic, and, when combined with $\widetilde{\mathbf{A}}^{FF}$, produces fields that obey causality (see Sec. \ref{sec:contours}).  Additionally, $\widetilde{F}$ only needs one MC constant: $\sqrt{2\pi} \rho_0$.  Although the second term in Eq. \ref{eq:eq5a} accounts for the asymmetric MC vector potential in an ad-hoc fashion, this asymmetry flows directly from Eq. \ref{eq:form7} (Sec. \ref{sec:time}), and special cases of the $x_{\rm i}$ are derived.  Rather than requiring six raw MC numbers, the associated code relies on Eq. \ref{eq:form7}, and one MC parameter ($\sqrt{2\pi} \rho_0$).

\subsubsection{Generalization of Eq. \ref{eq:form7}}
\label{sec:form3}

In the Appendix Sec. \ref{sec:form1}, the lateral distribution of excess charge near cascade maximum is shown to follow Eq. \ref{eq:form2} for $\rho' < \rho_{1}$, where $\rho_{1}$ is the Moli\`{e}re radius.  To include the effect of charges beyond a single Moli\`{e}re radius, the following form for $f(\mathbf{x}')$ may be taken:

\begin{equation}
f(\mathbf{x}') = \delta(z')\sum_{\rm i}^{N} a_{\rm i} \exp(-\sqrt{2\pi} \rho_{\rm i}\rho')
\label{eq:form8}
\end{equation}

The normalization requirement for the ICD provides the following constraint on the $2N$ free parameters:

\begin{equation}
\sum_{\rm i}^{N} \left( \frac{a_{\rm i}}{\rho_{\rm i}^2} \right) = 2 \pi
\label{eq:form9}
\end{equation}

Note that the units of the $a_{\rm i}$ parameters are the same as the normalization $\rho_{\rm 0}^2$ in the single-exponential case.  Let $\alpha_{\rm i}$ and $\sigma_{\rm i}$ take the following definitions:

\begin{align}
a_{\rm i} &= \alpha_{\rm i} \rho_{\rm i}^2
\label{eq:form10}
\\
\sigma_{\rm i} &= \frac{\gamma}{\sqrt{2\pi}\rho_{\rm i}}
\label{eq:form11}
\end{align}

With this definition, Eq. \ref{eq:form7} may be generalized to arbitrary Moli\`{e}re radii, taking the following form:

\begin{equation}
\widetilde{F}(\omega,\theta) = \sum_{\rm i}^{\rm N} \frac{\alpha_{\rm i}}{(1+\sigma_{\rm i}^2)^{3/2}}
\label{eq:form12}
\end{equation}

It is shown in Sec. \ref{sec:time2} that in the far-field limit, at $\theta = \theta_{\rm C}$, the effect of extending the form factor $\widetilde{F}$ to arbitrary Moli\`{e}re radii is equivalent to adding a set of additional poles to the Askaryan field in the complex $\omega$-plane.  In the time domain, the Askaryan field picks up a series of exponential terms corresponding to the added poles.

\subsection{Results of the Model: RB+LPM+$\widetilde{F}(\omega,\theta)$}
\label{sec:contours}

\begin{figure}
\centering
\begin{subfigure}{0.33\textwidth}
\centering
\includegraphics[width=\textwidth,trim=0cm 5cm 0cm 6cm,clip=true]{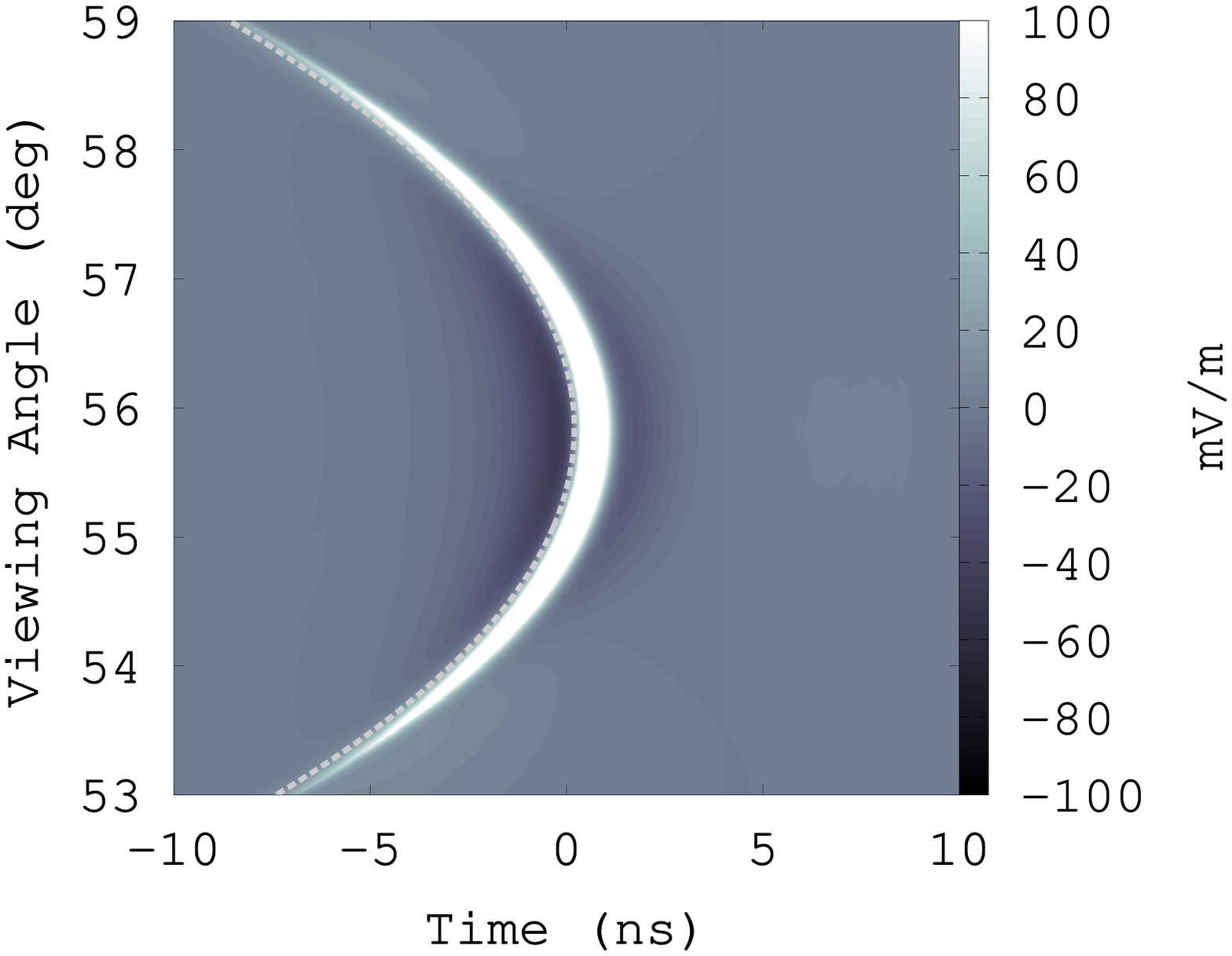}
\caption{}
\end{subfigure}%
\begin{subfigure}{0.33\textwidth}
\centering
\includegraphics[width=\textwidth,trim=0cm 5cm 0cm 6cm,clip=true]{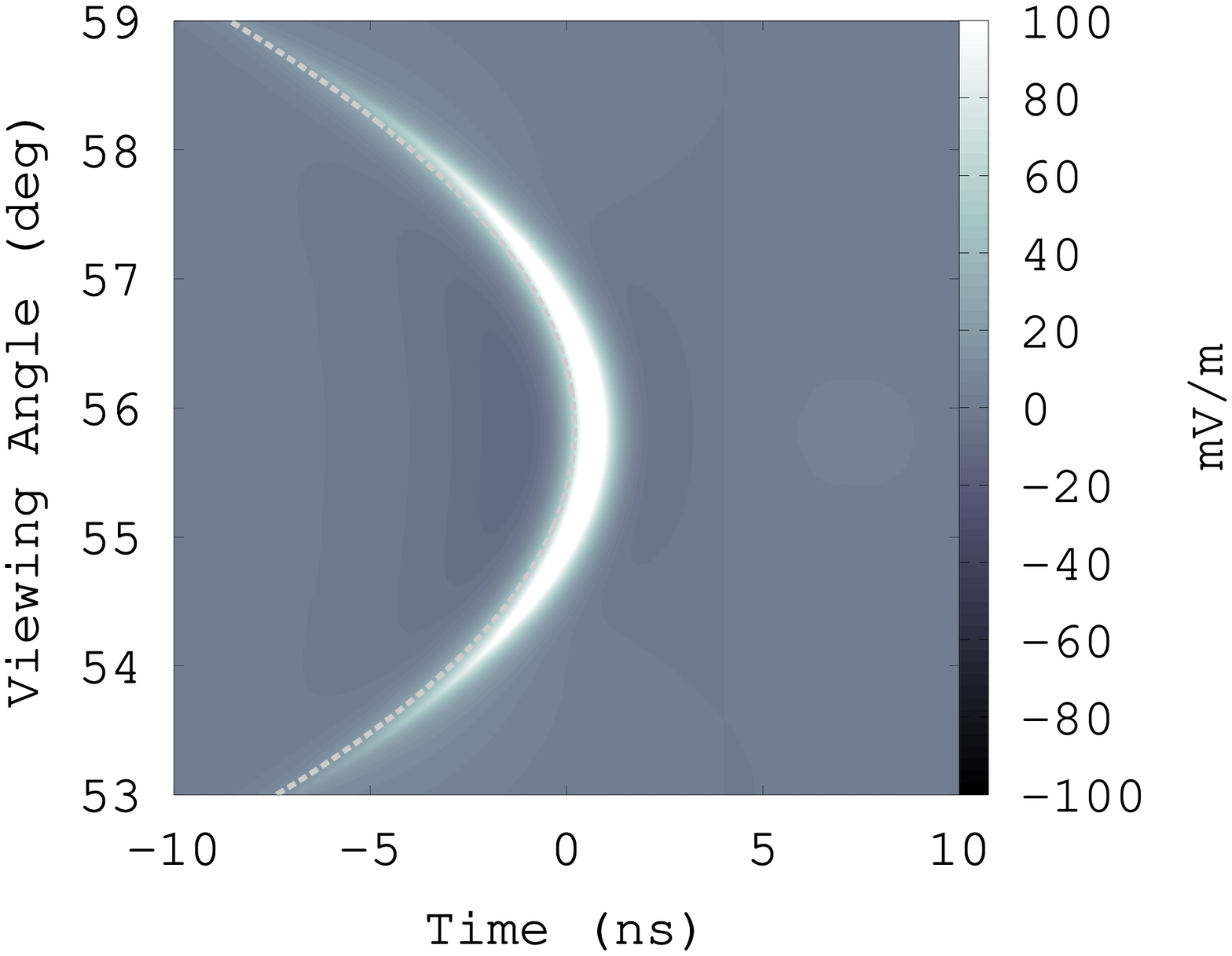}
\caption{}
\end{subfigure}
\centering
\begin{subfigure}{0.33\textwidth}
\centering
\includegraphics[width=\textwidth,trim=0cm 5cm 0cm 6cm,clip=true]{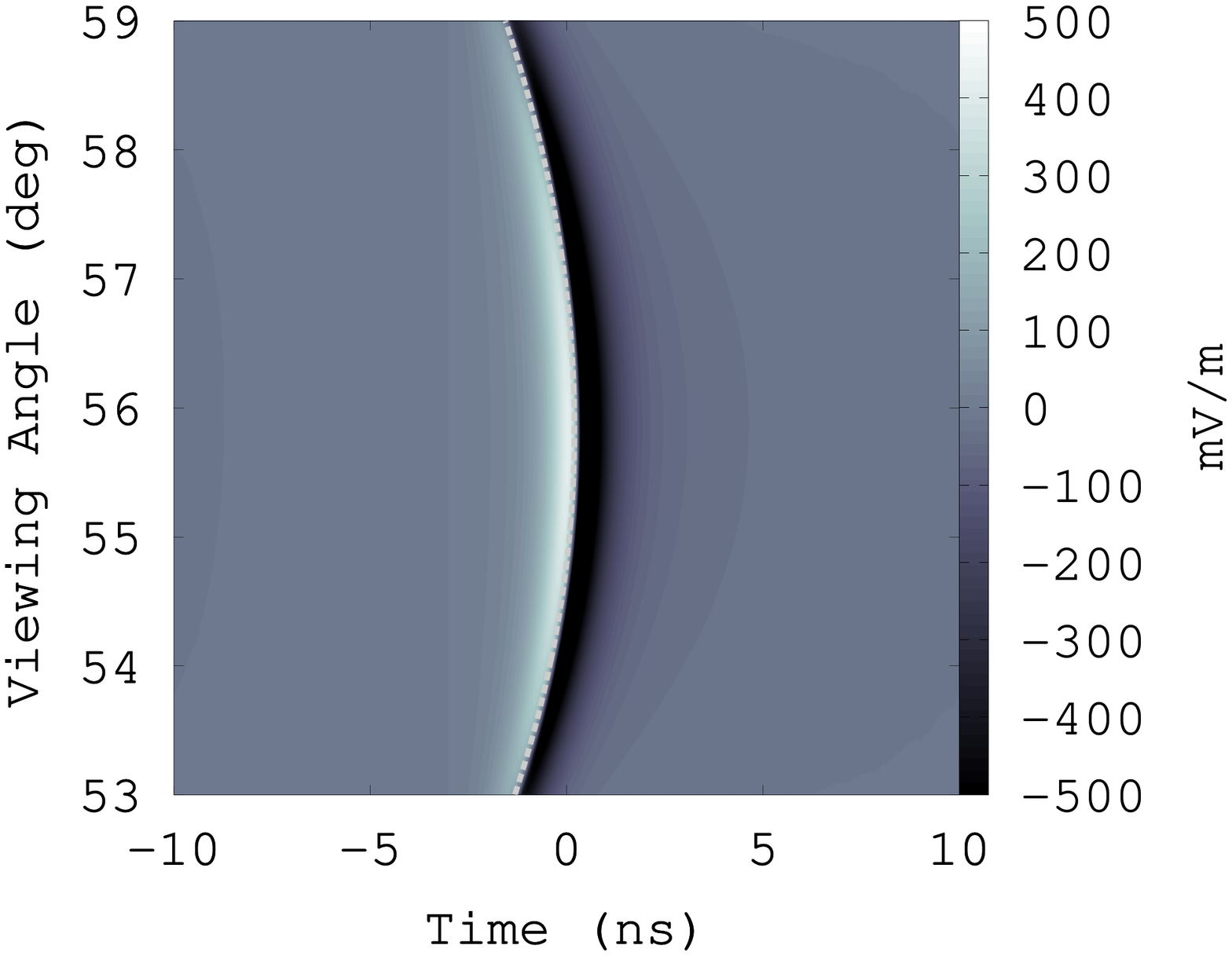}
\caption{}
\end{subfigure}%
\begin{subfigure}{0.33\textwidth}
\centering
\includegraphics[width=\textwidth,trim=0cm 5cm 0cm 6cm,clip=true]{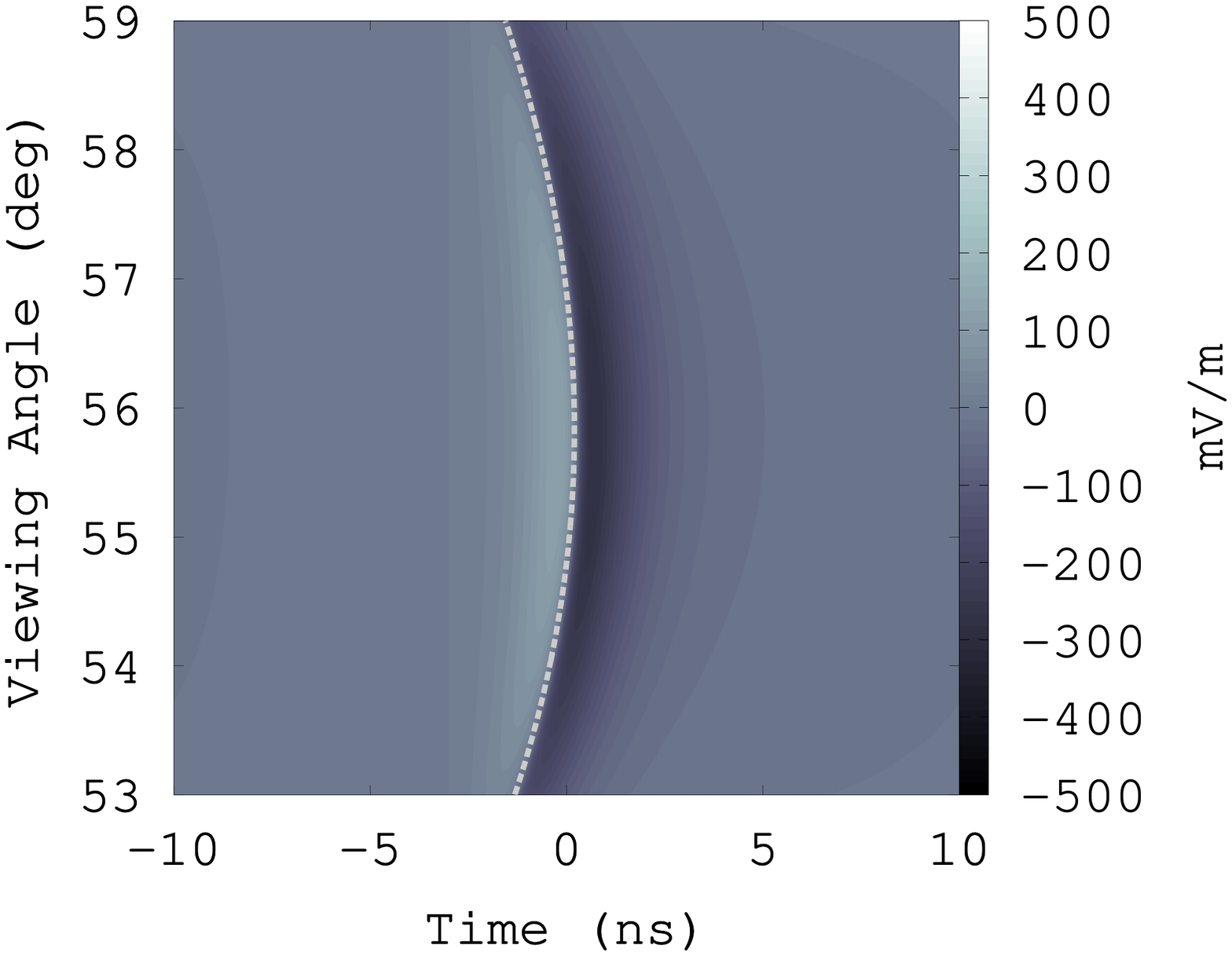}
\caption{}
\end{subfigure}
\caption{\label{fig:contours1} Contours of $\hat{e}_{\rm \theta} \cdot \mathbf{E}(t)$, for a cascade energy of 1000 PeV.  (a) R=1000 m, lateral ICD width of 5 cm.  (b) R=1000 m, lateral ICD width of 10 cm.  (c) R=200 m, lateral ICD width of 5 cm. (d) R=200 m, lateral ICD width of 10 cm.  The LPM effect has been taken into account.  See text for details.}
\end{figure}

The associated code field $\hat{e}_{\rm \theta} \cdot \mathbf{E}(t)$, including all effects, is shown in Fig. \ref{fig:contours1}, with $E_{\rm C} = 1000$ PeV.  Figure \ref{fig:contours1} contains contour graphs, in units of mV/m, versus the retarded time in nanoseconds, and $\theta$ in degrees.  The quadratic grey dashed line on the contours is a causal requirement from RB, showing how the arrival time (e.g. group delay) of the signal depends on $\theta$.  \textit{Phase delays} $t_{\rm \phi}$ about the quadratic are allowed: $t_{\rm \phi} = -\phi(\omega)/\omega$.  Phase delays are most prominent when $\widetilde{F}\neq 1$, $\theta \neq \theta_{\rm C}$, and when the LPM effect is strong.  See Appendix Sec. \ref{sec:rb1} for further detail.

The fields are shown for $R =$ 1000 m (panels a and b) and 200 m (panels c and d), $(\sqrt{2\pi}\rho_0)^{-1} =$ 5 cm (panels a and c) and 10 cm (panels b and d) in Fig. \ref{fig:contours1}.  The causality requirement from RB leads to off-cone regions have a higher effective velocity.  A larger $R$ value leads to wider separation in arrival times, as these off-cone modes have longer to outpace the other modes (earlier times correspond to more negative retarded times).  The larger $(\sqrt{2\pi}\rho_0)^{-1}$ values correspond to form factors that smear the fields, relative to smaller $(\sqrt{2\pi}\rho_0)^{-1}$ values.

An enticing implication of the effective velocity variation is that the degeneracy between a low-energy event interacting close to the observer, and a high-energy event interacting correspondingly farther from the observer would be broken.  Recall that $\widetilde{\textbf{E}} \propto R^{-1}$ in the far-field.  An event with $R = 100$ m and $E_{\rm C} = 10$ PeV would have the same amplitude as an event with $R = 1000$ m and $E_{\rm C} = 100$ PeV, neglecting secondary effects like ice absorption.  The temporal signature shown by the quadratics in Fig. \ref{fig:contours1} would be different in the two cases.

From Eq. \ref{eq:money3}, the field $\mathbf{\widetilde{E}}$ has both $\hat{e}_{\rm r}$ and $\hat{e}_{\rm \theta}$ components.  For the extreme Fraunhofer limit, as $\eta \rightarrow 0$, the ratio of the amplitudes of these components is independent of frequency:

\begin{equation}
\label{eq:comp1}
\frac{\hat{e}_{\rm r} \cdot \mathbf{\widetilde{E}}}{\hat{e}_{\rm \theta} \cdot \mathbf{\widetilde{E}}} = - \left( \frac{\cos\theta - \cos\theta_{\rm C}}{\sin\theta}\right)
\end{equation}

\begin{figure}
\centering
\begin{subfigure}{0.33\textwidth}
\centering
\includegraphics[width=\textwidth,trim=0cm 5cm 0cm 6cm,clip=true]{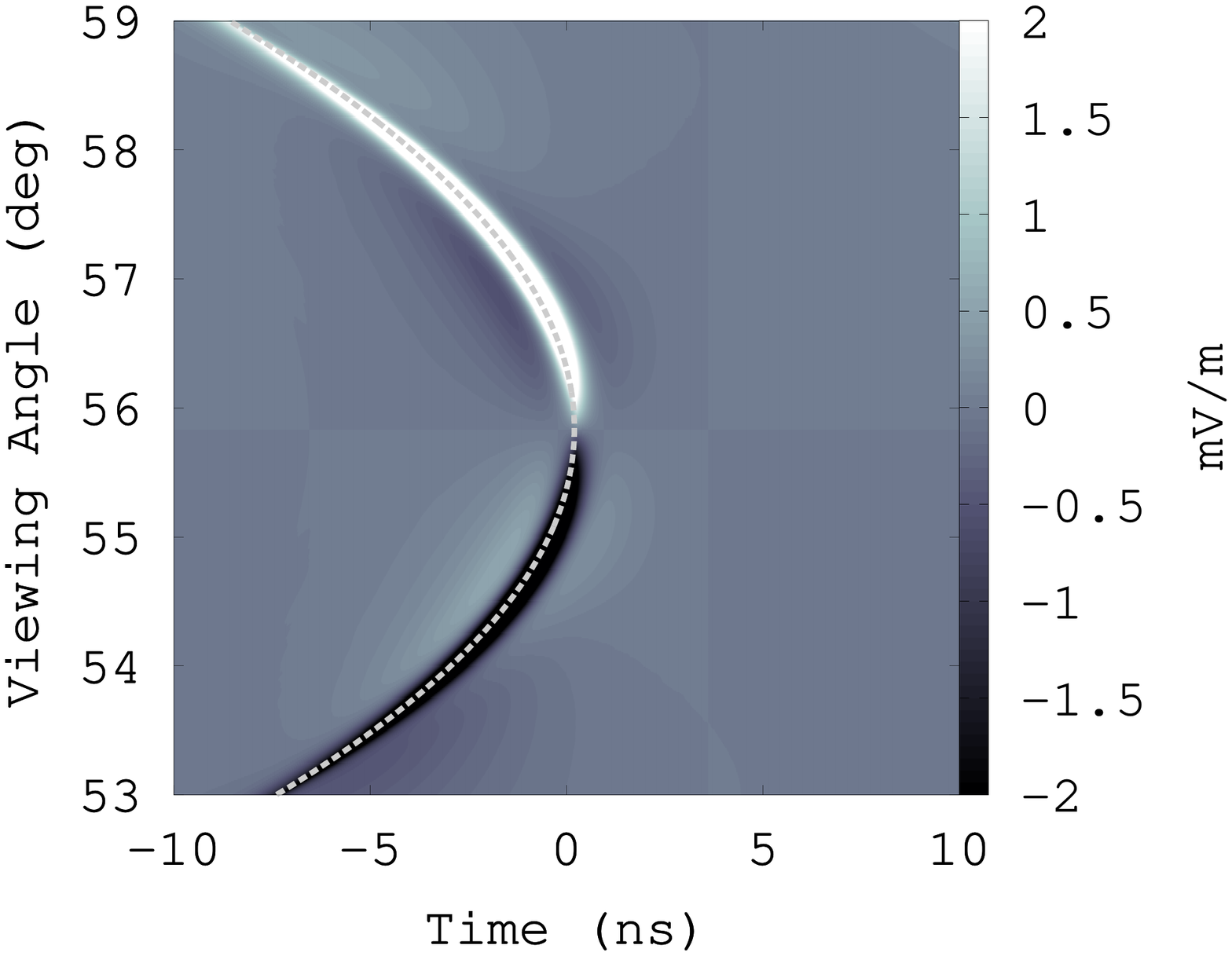}
\caption{}
\end{subfigure}%
\begin{subfigure}{0.33\textwidth}
\centering
\includegraphics[width=\textwidth,trim=0cm 5cm 0cm 6cm,clip=true]{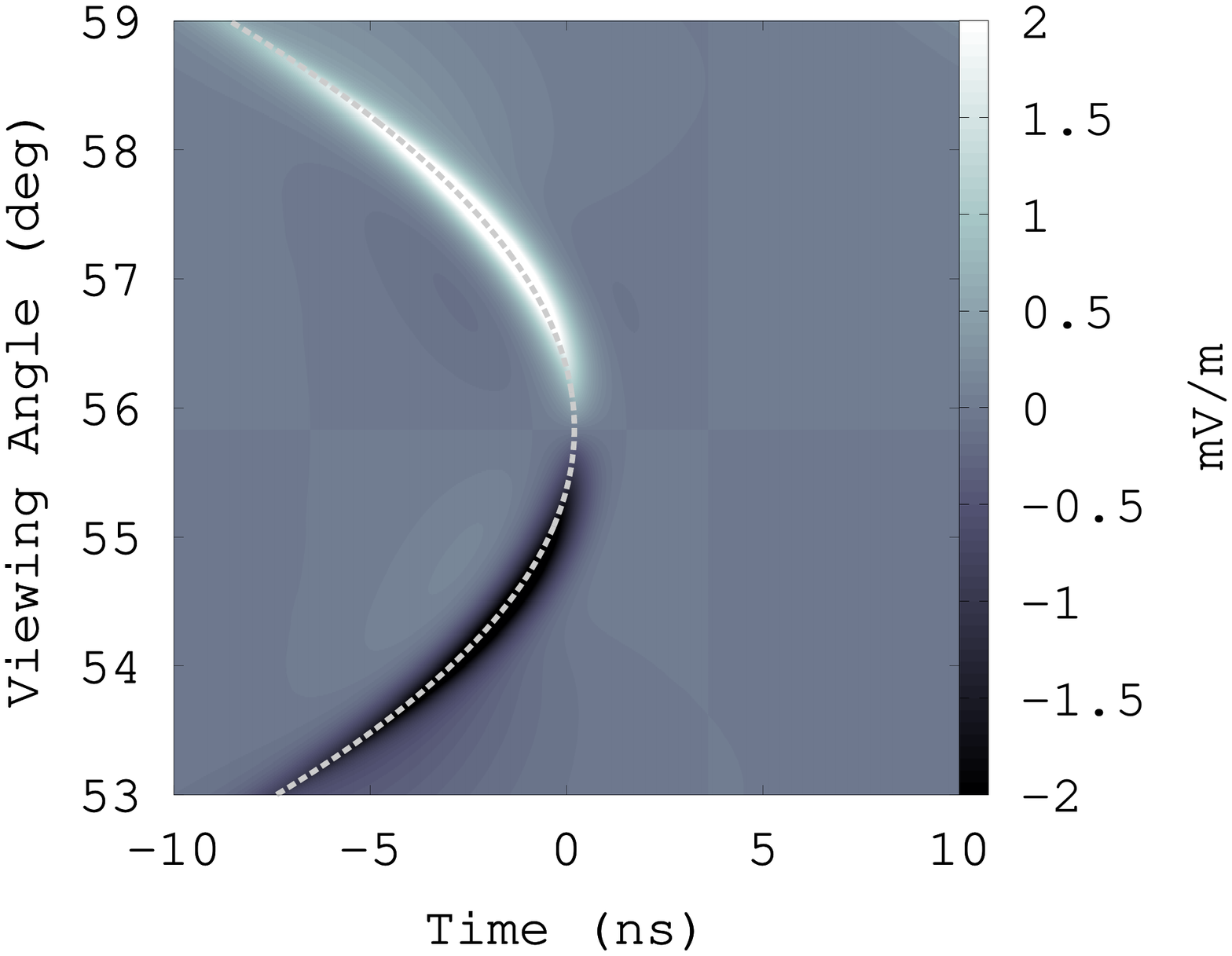}
\caption{}
\end{subfigure}
\centering
\begin{subfigure}{0.33\textwidth}
\centering
\includegraphics[width=\textwidth,trim=0cm 5cm 0cm 6cm,clip=true]{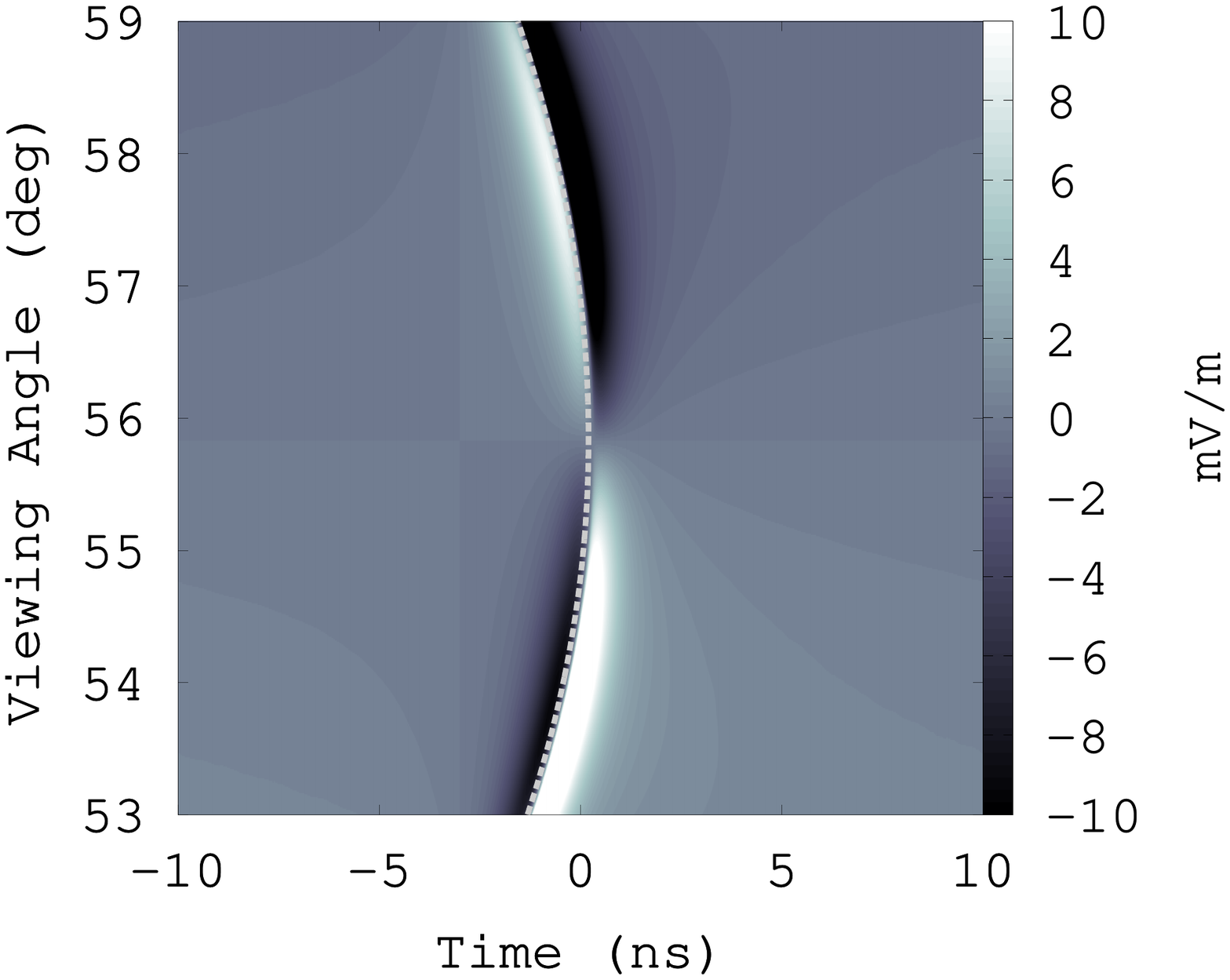}
\caption{}
\end{subfigure}%
\begin{subfigure}{0.33\textwidth}
\centering
\includegraphics[width=\textwidth,trim=0cm 5cm 0cm 6cm,clip=true]{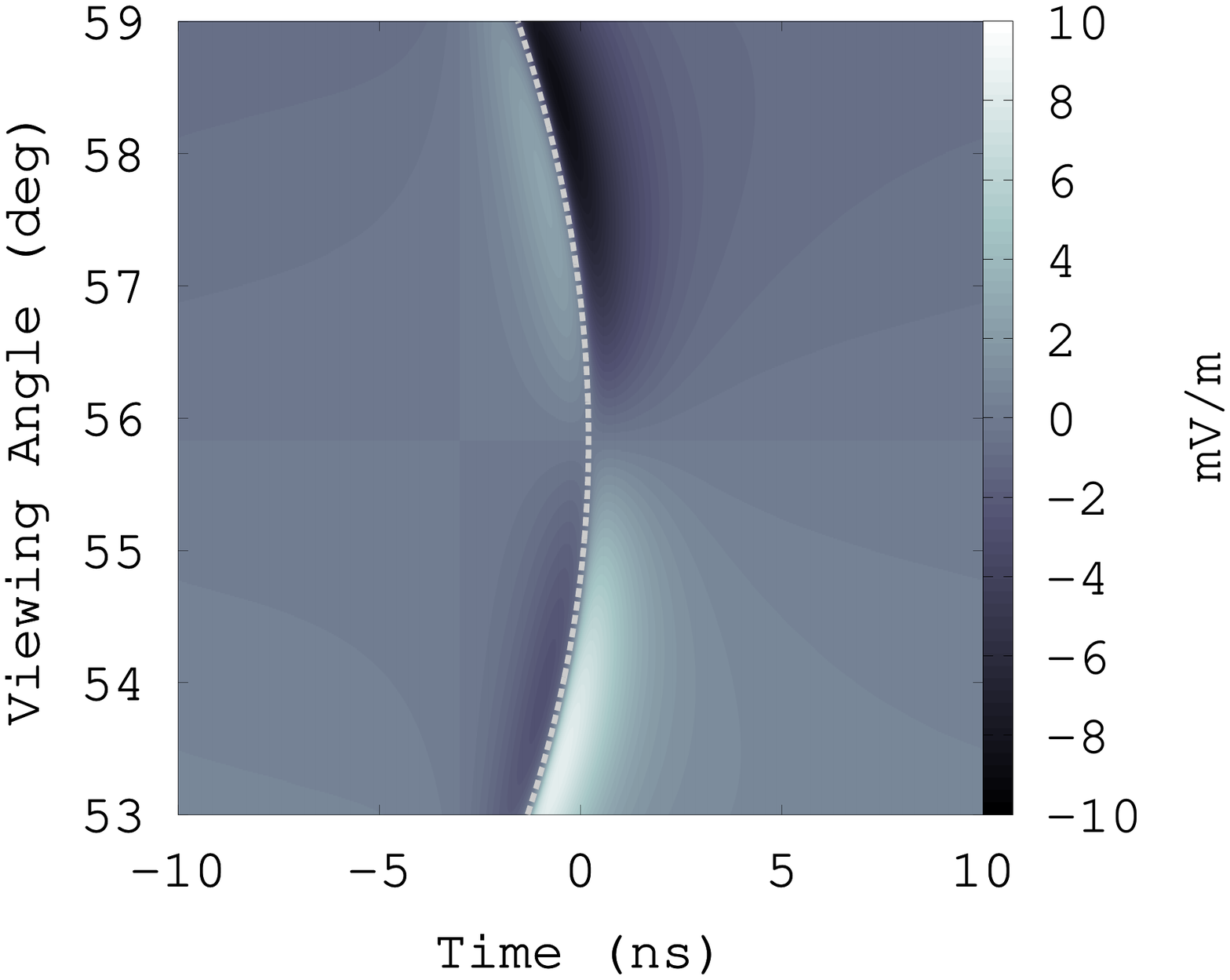}
\caption{}
\end{subfigure}
\caption{\label{fig:contours3} Contours of $\hat{e}_{\rm r} \cdot \mathbf{E}(t)$, for a cascade energy of 1000 PeV.  (a) R=1000 m, lateral ICD width of 5 cm.  (b) R=1000 m, lateral ICD width of 10 cm.  (c) R=200 m, lateral ICD width of 5 cm. (c) R=200 m, lateral ICD width of 10 cm.  In all cases, the gray dashed line represents the causality requirement.  See text for details.}
\end{figure}

Equation \ref{eq:comp1} shows that the $\hat{e}_{\rm r}$-component of $\mathbf{E}(t)$ is positive above the Cherenkov angle, and negative below it.  Since the $\hat{e}_{\rm r} \cdot \mathbf{E}(t) = 0$ at $\theta_C$, the maximum in the $\hat{e}_{\rm r}$-component is always at some angle $\theta \neq \theta_C$.  The contour graphs of Fig. \ref{fig:contours3} represent the $\hat{e}_{\rm r}$-component of the same fields as Fig. \ref{fig:contours1}.  Because $\hat{e}_{\rm r} \cdot \mathbf{\widetilde{E}} < \hat{e}_{\rm \theta} \cdot \mathbf{\widetilde{E}}$, the Askaryan field is usually given with a pure $\hat{e}_{\rm \theta}$-polarization.  Though the $\hat{e}_{\rm r}$-component is small compared to the $\hat{e}_{\rm \theta}$-component, the code does not neglect it.  The polarization ratio (Eq. \ref{eq:comp1}) is both complex, and frequency-dependent if $\eta \neq 0$.

\subsection{Spectral, Phase, and Angular Comparisons to Previous Work}
\label{sec:rb1}

\begin{figure}[ht]
\centering
\begin{subfigure}{0.5\textwidth}
\centering
\includegraphics[width=0.7\textwidth,angle=270]{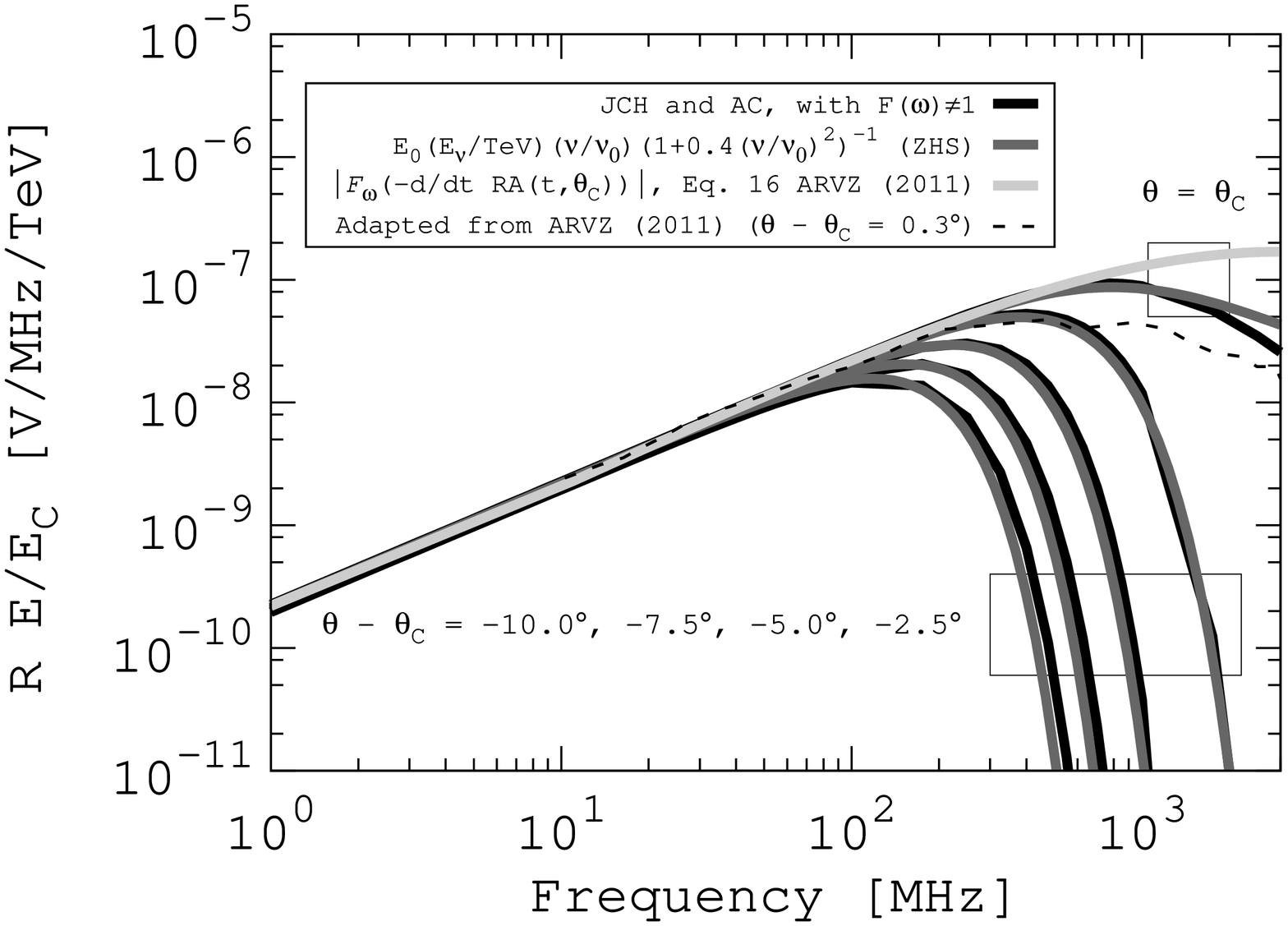}
\caption{}
\end{subfigure}%
\begin{subfigure}{0.5\textwidth}
\centering
\includegraphics[width=0.66\textwidth,angle=270]{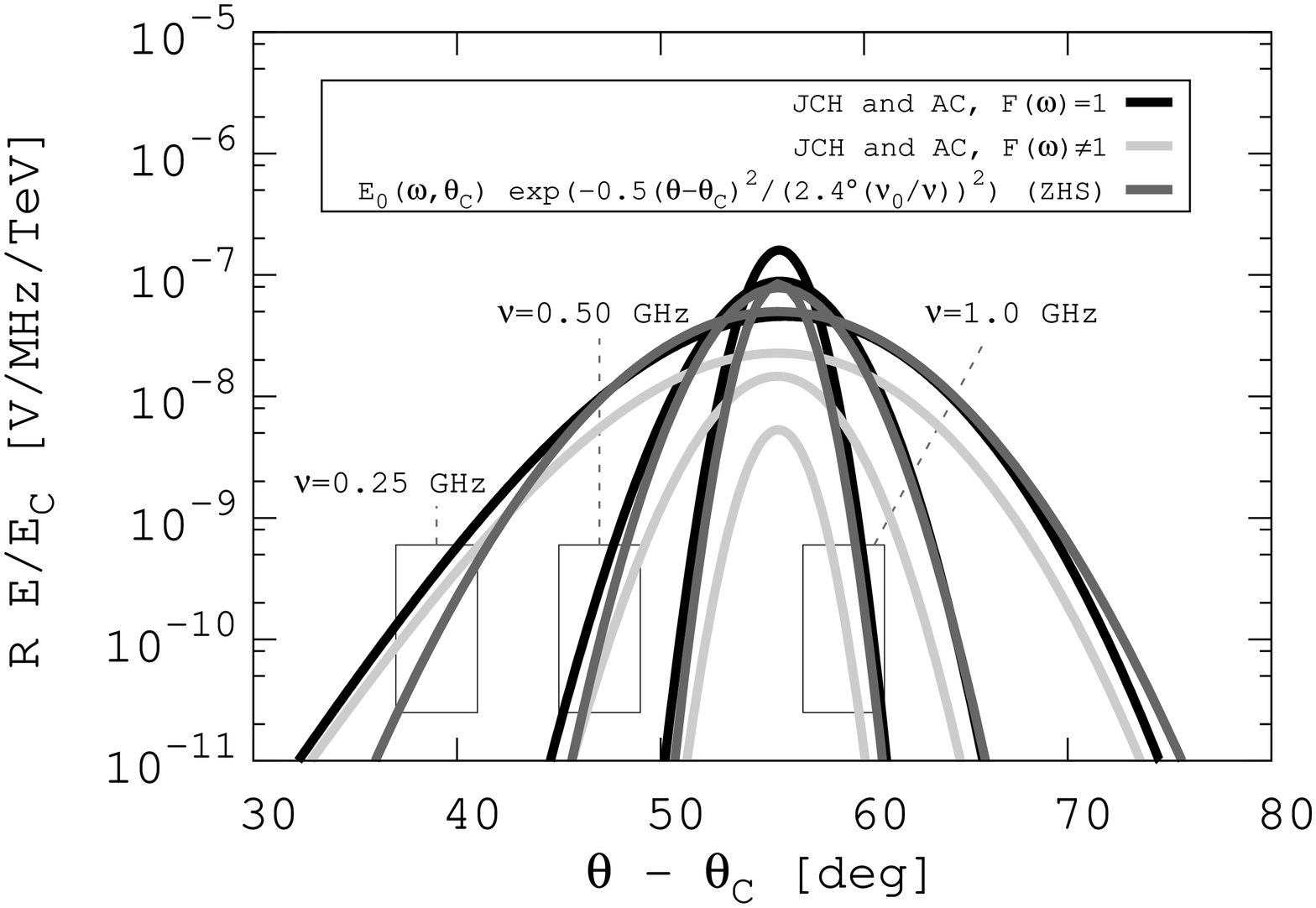}
\caption{}
\end{subfigure}
\caption{\label{fig:rb1} (a) A comparison of the phase of the Askaryan field in different models versus $\nu$ for $E_{\rm C} = 100$ TeV scaled by $R/E_{\rm C}$.  Models shown are: the associated code $|\hat{e}_{\rm \theta} \cdot \mathbf{\widetilde{E}}(\omega)|$ (black), ZHS parameterization (dark gray), the $|\hat{e}_{\rm \theta} \cdot \mathbf{\widetilde{E}}(\omega)|$ from Eq. 16 from ARVZ \cite{PhysRevD.84.103003} (light gray), and the spectrum adapted from Fig. 3 of \cite{PhysRevD.84.103003} (dashed line).  For the associated code (black), $a=1.5$ m, $R = 1000$ m, $\widetilde{F} \neq 1$.  The upper right box encompasses cases for which $\theta=\theta_{\rm C}$, and the lower left box encompasses cases for which $\theta\neq\theta_{\rm C}$.  The cases are (from right to left) $\theta_{\rm C}-2.5^{\circ}$, $\theta_{\rm C}-5.0^{\circ}$, $\theta_{\rm C}-7.5^{\circ}$, and $\theta_{\rm C}-10.0^{\circ}$.  (b) The angular dependence of the associated code $|\hat{e}_{\rm \theta} \cdot \mathbf{\widetilde{E}}(\omega)|$, $\widetilde{F}(\omega)=1$ (black), ZHS parameterization (dark gray), and the associated code $|\hat{e}_{\rm \theta} \cdot \mathbf{\widetilde{E}}(\omega)|$, $\widetilde{F}(\omega)\neq1$ (dashed).}
\end{figure}

\begin{figure}
\begin{center}
\includegraphics[width=0.33\textwidth,angle=270]{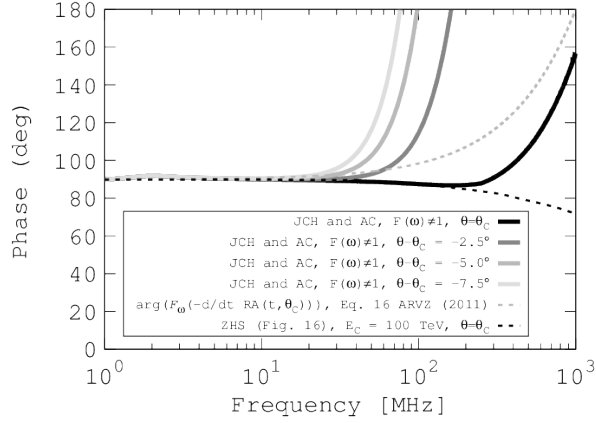}
\end{center}
\caption{\label{fig:phase}  A comparison of the phase of the Askaryan field in different models versus $\nu$ for $E_{\rm C} = 100$ TeV, $R=1000$ m.  The associated code produces the solid lines at viewing angles of $\theta_{\rm C}$ and $\theta-\theta_{\rm C} =\lbrace -2.5^{\circ}, -5.0^{\circ}, -7.5^{\circ}\rbrace$.  The gray dashed line is the phase of Eq. 16 of \cite{PhysRevD.84.103003} ($\theta=\theta_{\rm C}$), and the black dashed line is the ZHS result.}
\end{figure}

It is shown in Figs. \ref{fig:rb1} and \ref{fig:phase} where the associated code agrees with the ZHS parameterization for $E_{\rm C} < E_{\rm LPM}$.  Equations \ref{eq:money}-\ref{eq:money4}, via the associated code, are compared in Figs. \ref{fig:rb1}-\ref{fig:phase} to Eqs. 20-21 from \cite{PhysRevD.45.362}, with $\Delta \theta = 2.4^{\circ} (\nu_{\rm 0}/\nu)$, and $\nu_{\rm 0}=0.5$ GHz, shown for convenience in Eqs. \ref{eq:zhs1}-\ref{eq:zhs2}.  In Fig. \ref{fig:rb1}, the spectra are scaled by $R$[m]/$E_{\rm C}$[TeV], where $E_{\rm C}$ is the cascade energy in TeV.

\begin{align}
\label{eq:zhs1}
\frac{R|\mathbf{\widetilde{E}}(\omega,\theta=\theta_{\rm C})|}{\left[\frac{\SI{}{\V}}{\SI{}{\MHz}}\right]} &= 1.1\times 10^{-7} \frac{E_{\rm 0}}{[\SI{}{\TeV}]} \left(\frac{\nu}{\nu_{\rm 0}}\right) \frac{1}{1+(\nu/\nu_{\rm 0})^2} \hat{e}_{\rm \theta} \\
\label{eq:zhs2}
\mathbf{\widetilde{E}}(\omega,\theta) &= \mathbf{\widetilde{E}}(\omega,\theta=\theta_{\rm C}) \exp\left[-\frac{1}{2}\left(\frac{\theta-\theta_{\rm C}}{\Delta \theta}\right)^2\right] \hat{e}_{\rm \theta}
\end{align}

The $|\mathbf{\widetilde{E}}(\omega,\theta_{\rm C})|$ derived from Eq. 16 from ARVZ \cite{PhysRevD.84.103003} is also shown in Fig. \ref{fig:rb1} (a) for comparison, as well as an off-cone example given in Fig. 3 of ARVZ \cite{PhysRevD.84.103003}.  The on-cone ARVZ case is given as a vector potential, so $\textbf{E}(t) = - \dot{\textbf{A}}(t,\theta_{\rm C})$ is applied before extracting $|\mathbf{\widetilde{E}}(\omega,\theta_{\rm C})|$.  When $\theta=\theta_{\rm C}$, the various $\widetilde{F}(\omega)$ are the cause of the attenuation above 1 GHz.  The choice of $\widetilde{F}(\omega)$ for the associated code corresponds to a lateral cascade width of $\approx 5$ cm, and $\widetilde{F}(\omega)$ attenuates the spectrum above 1 GHz more realistically than ZHS.  When $\theta\neq\theta_{\rm C}$, the width of the Cherenkov cone is the cause of the high-frequency attenuation.  The associated code is compared to ZHS at several angles, and the off-cone ARVZ lies in between the other results above 1 GHz, due to the smaller off-cone angle (0.3$^{\circ}$).  In Fig. \ref{fig:rb1} (b), the angular dependence of the associated code and ZHS is compared at 0.25, 0.5 and 1.0 GHz.  The cone-width is also inversely proportional to $a$ in the RB model - an important feature that accounts for cone-width narrowing when LPM elongation is important (see Figs. \ref{fig:lpm}-\ref{fig:lpm2} below).  In Fig. \ref{fig:rb1} (b), the results with $\widetilde{F}\neq 1$ are systematically lower than those with $\widetilde{F}=1$, as expected.

The phase $\phi(\omega)$ of $\hat{e}_{\rm \theta} \cdot \mathbf{\widetilde{E}}$ versus frequency is shown in Fig. \ref{fig:phase}, for several values of $\theta-\theta_{\rm C}$.  The phase $\phi(\omega)$ is unwrapped after taking the arctangent of the ratio of the imaginary to the real parts of the Fourier transform of the model waveforms.  Shown for comparison are the phases of Eq. 16 of ARVZ \cite{PhysRevD.84.103003}, and ZHS (Fig. 16 of \cite{PhysRevD.45.362}), both at $\theta=\theta_{\rm C}$.  For the ARVZ case, which is a vector potential $\textbf{A}(t,\theta_{\rm C})$, $\textbf{E}(t) = - \dot{\textbf{A}}(t,\theta_{\rm C})$ is applied before extracting the phase.  All the models have $\phi(\omega) \sim 90^{\circ}$ for $\nu < 100$ MHz \footnote{$\mathcal{F}_{\rm \omega}(- \dot{\textbf{A}}(t,\omega_{\rm C})) = \widetilde{\textbf{E}}(\omega,\theta_{\rm C}) = -i\omega\widetilde{\textbf{A}}(\omega,\theta_{\rm C}) = -e^{i\pi/2}\omega\widetilde{\textbf{A}}(\omega,\theta_{\rm C})$, so a low-frequency phase of $\pi/2$ (see Sec. \ref{sec:caus})}.  The $\phi(\omega)$ functions diverge above 100 MHz, where the radiated power is reduced off-cone.  The group delay ($-d\phi/d\omega$) in this regime is roughly constant.  Therefore, the ARVZ model and the associated code differ in phase above 100 MHz, but not group delay.  Exploring all effects on $\phi(\omega)$, including LPM elongation and coherence, is outside the scope of this work.

The influence of LPM elongation on $|\hat{e}_{\rm \theta} \cdot \mathbf{\widetilde{E}}(\omega)|$ is shown in Fig. \ref{fig:lpm}, revealing two effects relative to other models.  First, there is a mild enhancement below 100 MHz in panels (b) and (d).  Figure \ref{fig:lpm}, panels (a) and (c) show spectra with the \textit{strictLowFreqLimit} flag activated in the associated code (see discussion in Sec. \ref{sec:lpm}).  Figure \ref{fig:lpm}, panels (b) and (d) show spectra without the \textit{strictLowFreqLimit} flag, by default.  As long as $\eta = ka^2\sin^2\theta/R \lesssim 1$, so that the stretched $a$ is still $\lesssim \Delta z_{\rm coh}$, the mild low-frequency enhancement in panels (b) and (d) may be attributed to radiation from an elongated cascade.

Second, spectra in panels (a) and (b) have different values of $\eta$ than the spectra in panels (c) and (d), leading to a counter-intuitive high-frequency dependence.  Equation \ref{eq:money4} at fixed $\theta$ implies a Gaussian frequency distribution: $|\mathcal{W}(\eta,\theta)| \propto \exp(-\frac{1}{2}(\Delta\cos\theta)^2(ka)^2)/(1+\eta^2)$, with $\Delta\cos\theta = \cos\theta-\cos\theta_{\rm C}$.  The Gaussian width in $\nu$ is

\begin{equation}
\sigma_{\rm \nu}(a,\eta,\theta) = \frac{c}{2\pi a\Delta\cos\theta}\left(1+\eta^2\right)^{1/2}
\label{eq:width}
\end{equation}

For $a \lesssim R$, as in panels (a) and (b), $\eta$-term on the right side of Eq. \ref{eq:width} dominates for large $\nu$ because $\eta > 1$, which leads to $\sigma_{\rm \nu} \sim \nu (a/R)$.  The spectra in panels (a) and (b) are wider than the low-energy spectra in panels (c) and (d) due to the frequency-dependent $\sigma_{\rm \nu}$, which changes the high-frequency attenuation from Gaussian to linear attenuation at high-frequencies.  Conversely, the low-energy spectra in panels (c) and (d) have $a \ll R$, and $\eta \ll 1$, implying from Eq. \ref{eq:width} that $\sigma_{\rm \nu} \sim 1/a$, while the high-energy spectra encounter the former limit, because the $a$ values are enlarged by LPM elongation.  Thus, the low-energy spectra do not have the enhancement at high-frequencies, but the high-energy spectra have the enhancement.  These various limits are important because the simulated range of neutrino interaction ranges detectable to ARA/ARIANNA are typically 100-1000 m \cite{Allison2012457} \cite{DookaykaThesis}.

\begin{figure}[ht]
\centering
\begin{subfigure}{0.4\textwidth}
\centering
\includegraphics[width=0.7\textwidth,angle=270,trim=0cm 1.2cm 0cm 0cm,clip=true]{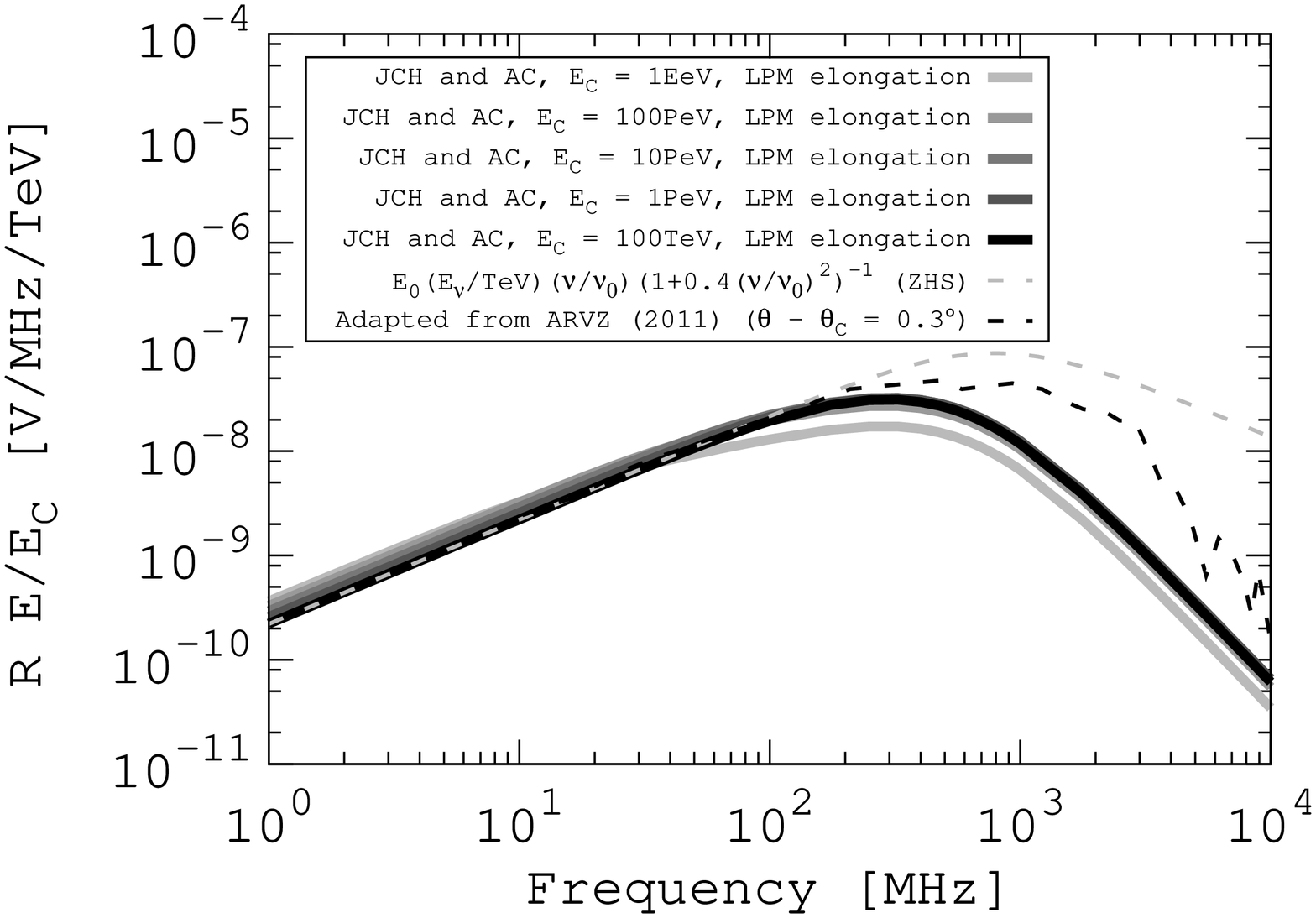}
\caption{}
\end{subfigure}%
\begin{subfigure}{0.4\textwidth}
\centering
\includegraphics[width=0.7\textwidth,angle=270,trim=0cm 1.2cm 0cm 0cm,clip=true]{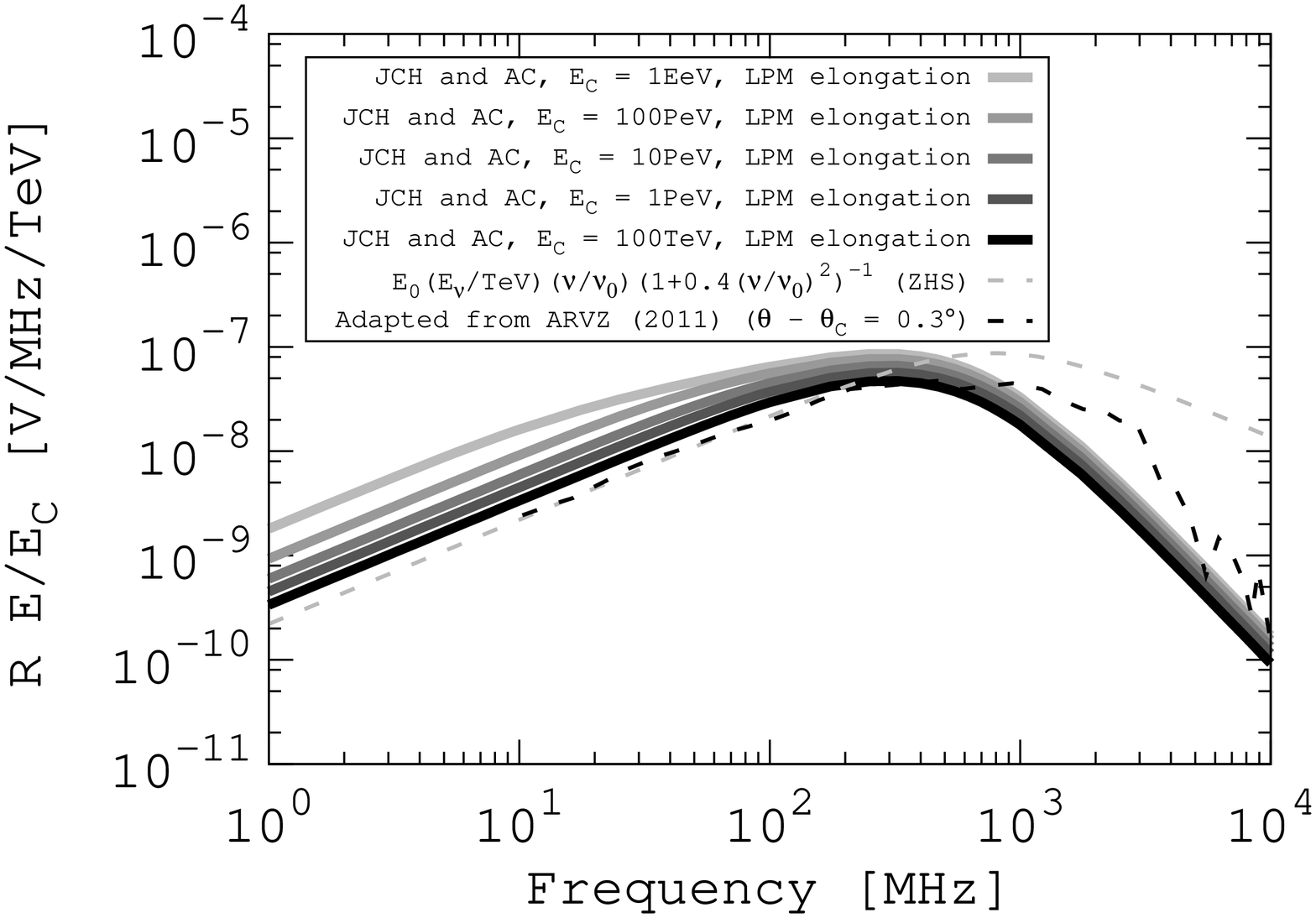}
\caption{}
\end{subfigure}
\centering
\begin{subfigure}{0.4\textwidth}
\centering
\includegraphics[width=0.7\textwidth,angle=270,trim=0cm 1.2cm 0cm 0cm,clip=true]{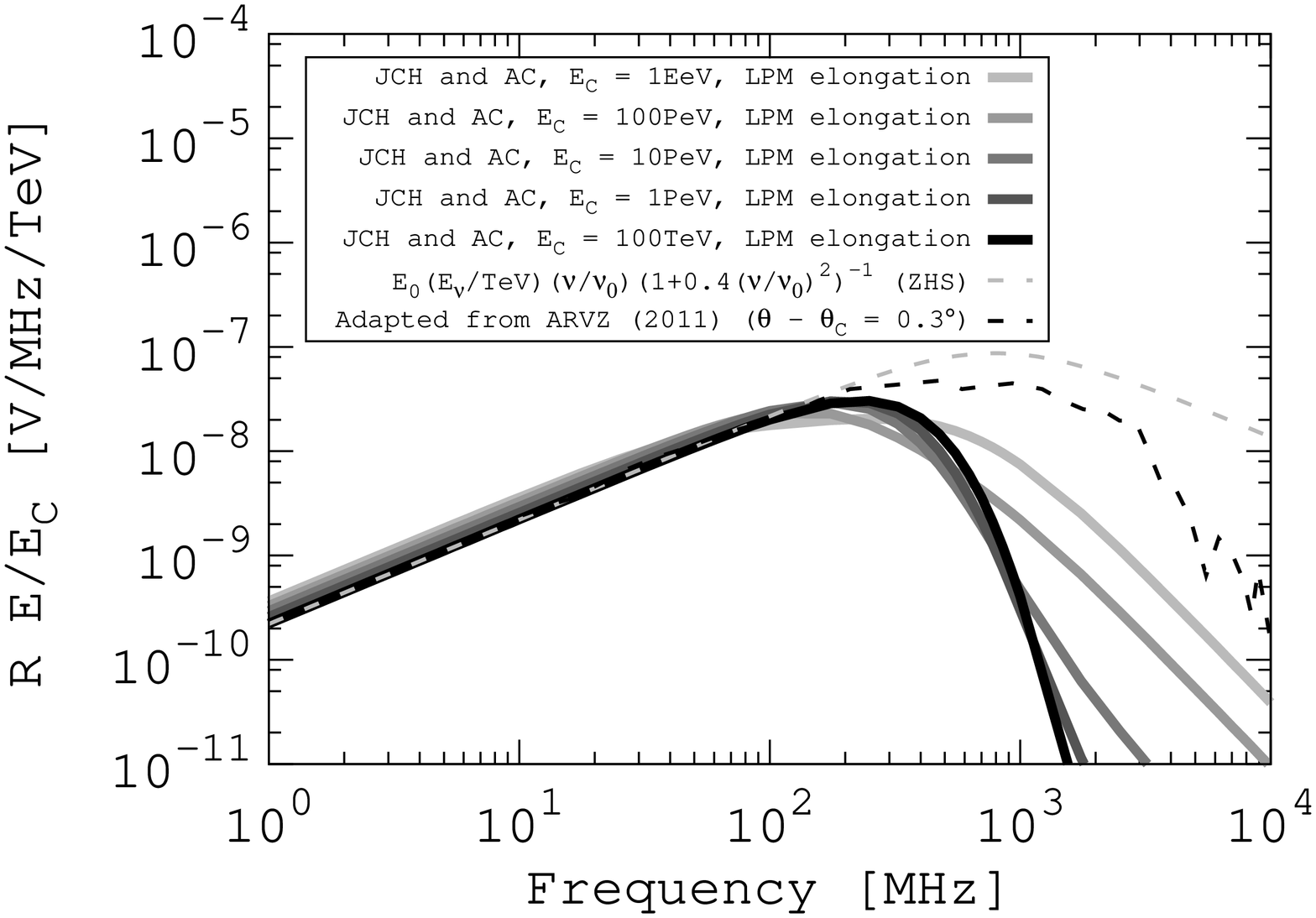}
\caption{}
\end{subfigure}%
\begin{subfigure}{0.4\textwidth}
\centering
\includegraphics[width=0.7\textwidth,angle=270,trim=0cm 1.2cm 0cm 0cm,clip=true]{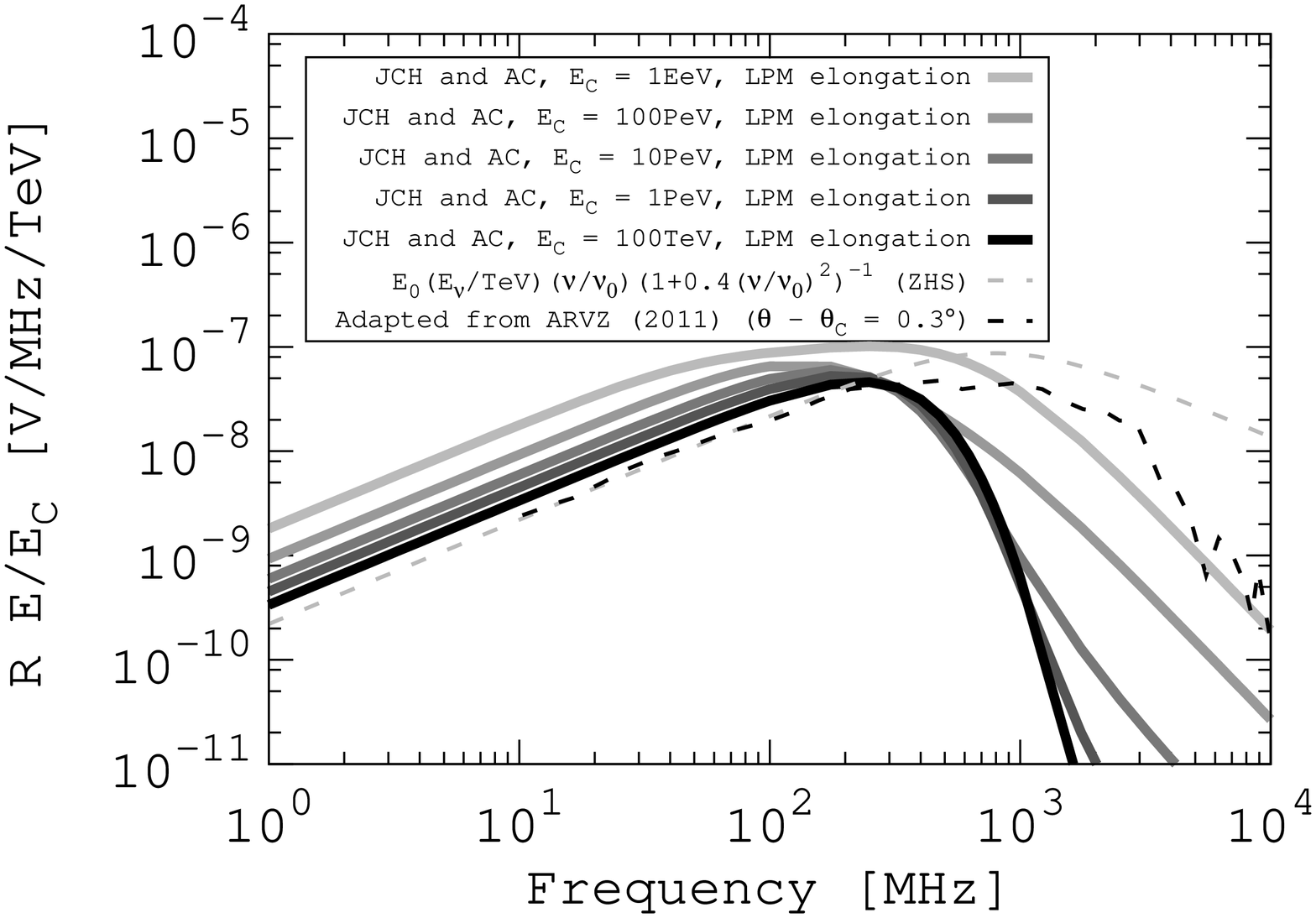}
\caption{}
\end{subfigure}
\caption{\label{fig:lpm} The influence of LPM elongation on $|\hat{e}_{\rm \theta} \cdot \mathbf{\widetilde{E}}(\omega)|$, for different values of $\eta$.  (a) $|\hat{e}_{\rm \theta} \cdot \mathbf{\widetilde{E}}(\omega)|$, for $R = 200$ m, and $\theta = 57^{\circ}$, low-frequency restricted.  (b) $|\hat{e}_{\rm \theta} \cdot \mathbf{\widetilde{E}}(\omega)|$, for $R = 200$ m, and $\theta = 57^{\circ}$.  (c) $|\hat{e}_{\rm \theta} \cdot \mathbf{\widetilde{E}}(\omega)|$, for $R = 1000$ m, and $\theta = 57^{\circ}$, low-frequency restricted.  (d) $|\hat{e}_{\rm \theta} \cdot \mathbf{\widetilde{E}}(\omega)|$, for $R = 1000$ m, and $\theta = 57^{\circ}$.  In all graphs, the black dashed line is the ZHS result with no LPM elongation, and the gray dashed line is the ARVZ result adapted from Fig. 3 of \cite{PhysRevD.84.103003}.  See text for detail.}
\end{figure}

\begin{figure}[ht]
\centering
\begin{subfigure}{0.4\textwidth}
\centering
\includegraphics[width=0.68\textwidth,angle=270]{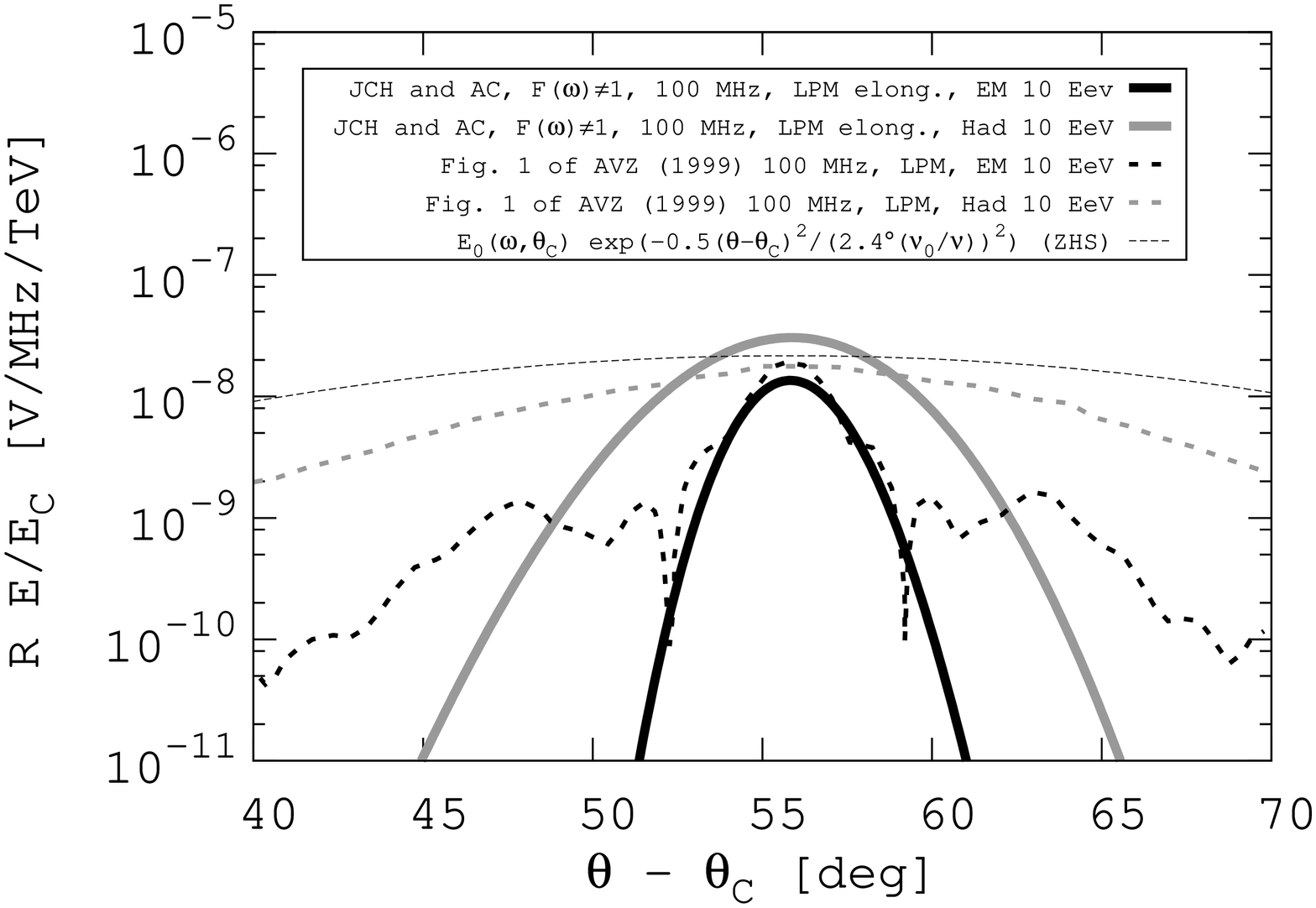}
\caption{}
\end{subfigure}%
\begin{subfigure}{0.4\textwidth}
\centering
\includegraphics[width=0.68\textwidth,angle=270]{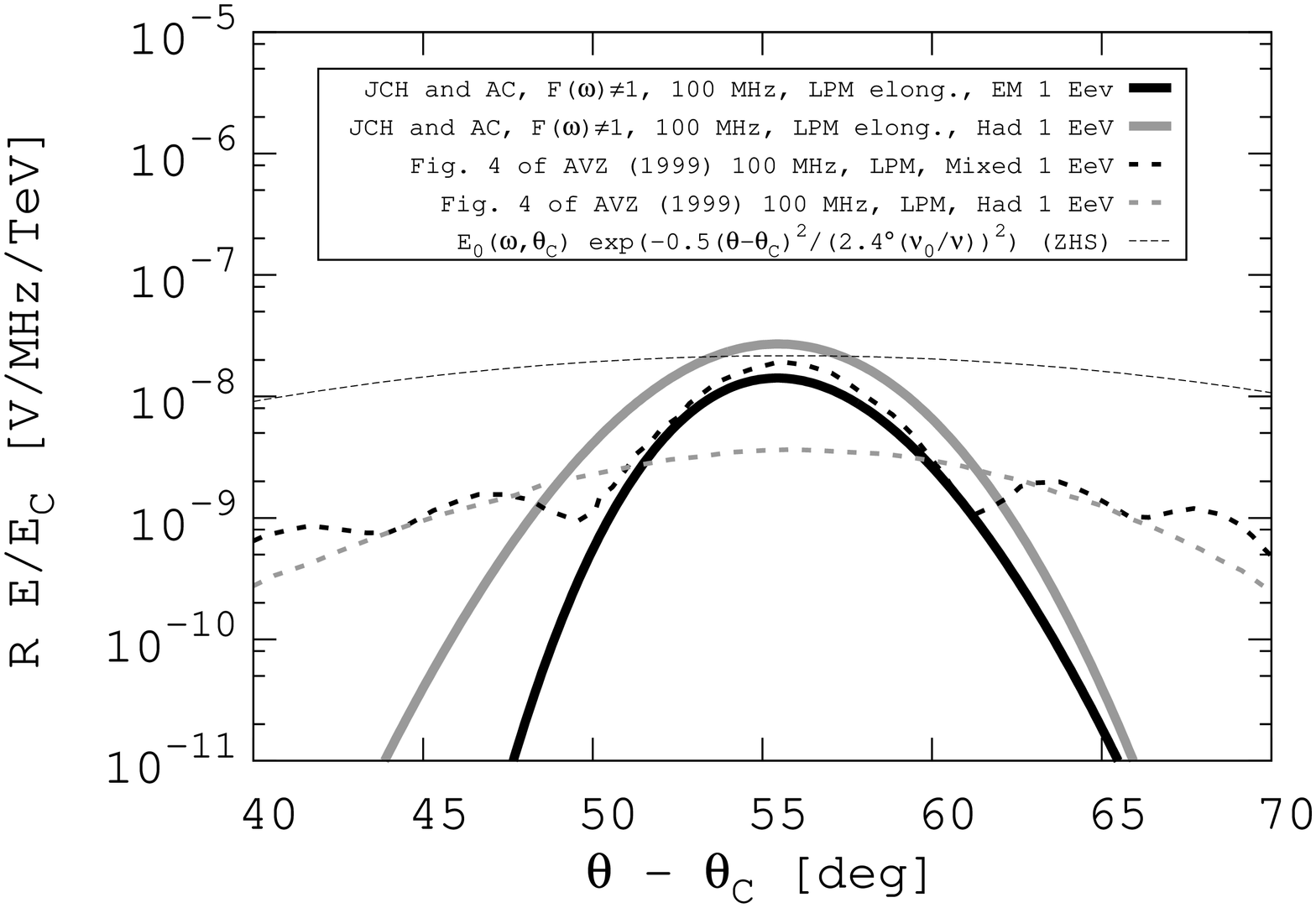}
\caption{}
\end{subfigure}
\centering
\begin{subfigure}{0.4\textwidth}
\centering
\includegraphics[width=0.68\textwidth,angle=270]{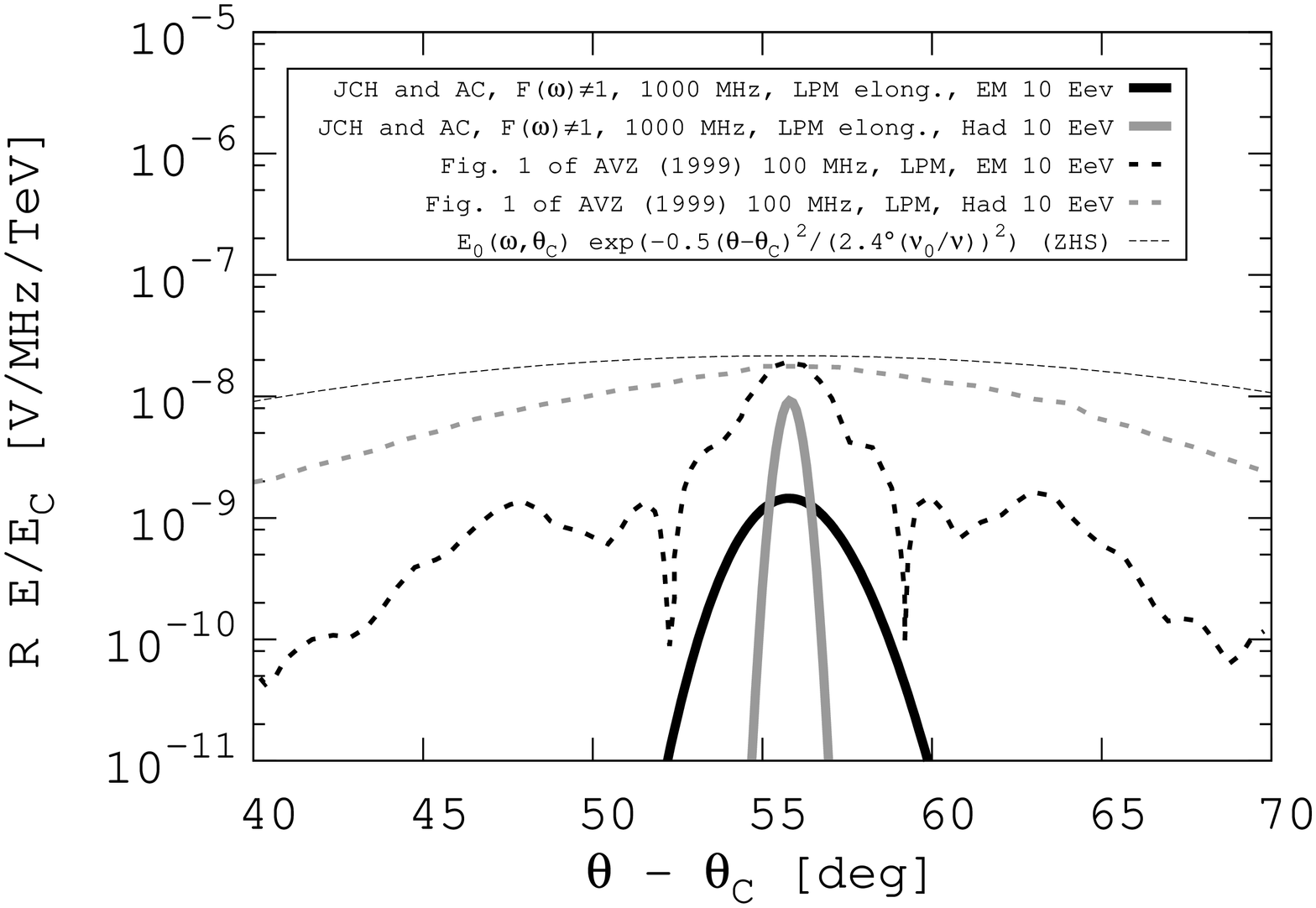}
\caption{}
\end{subfigure}%
\begin{subfigure}{0.4\textwidth}
\centering
\includegraphics[width=0.68\textwidth,angle=270]{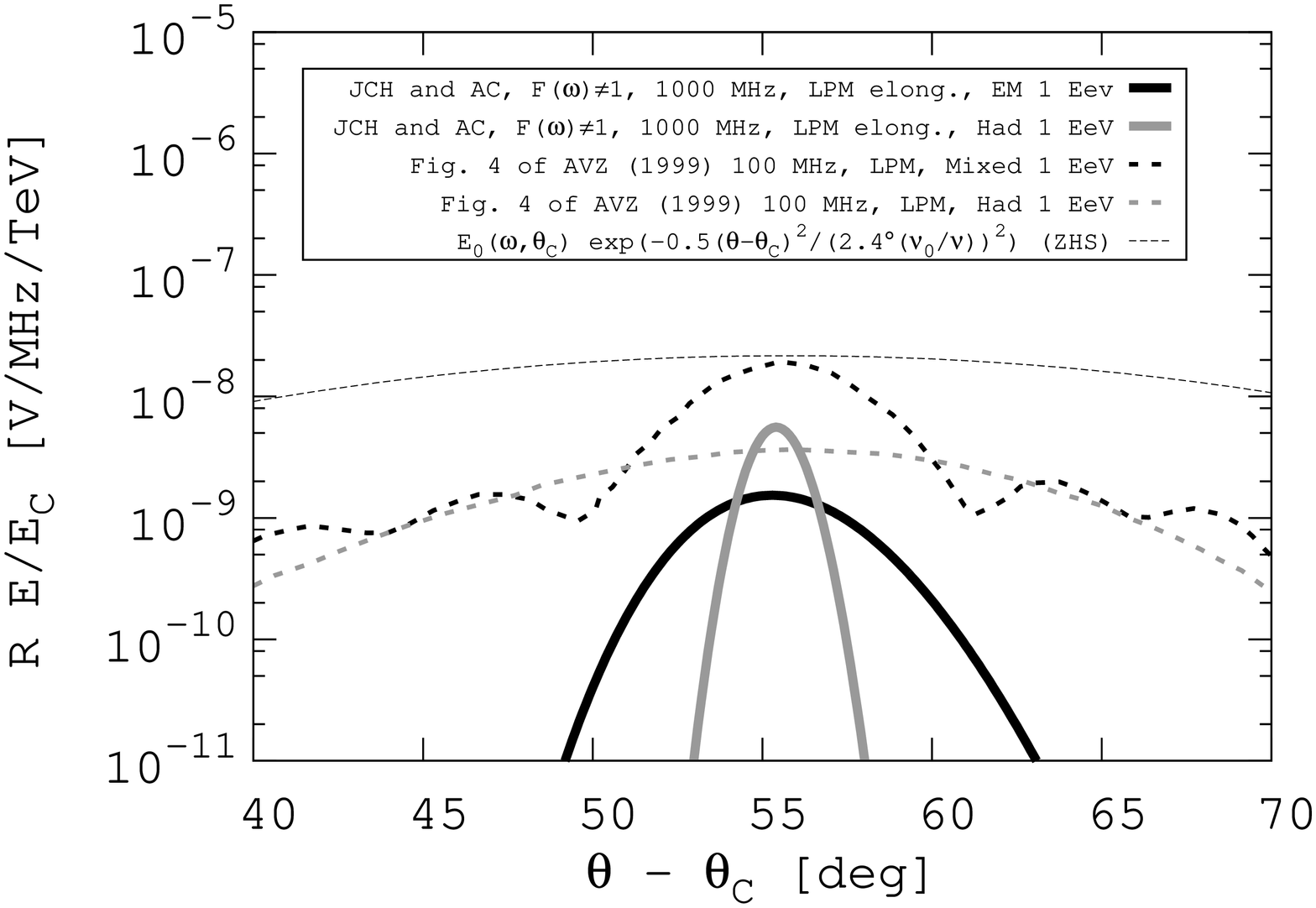}
\caption{}
\end{subfigure}
\caption{\label{fig:lpm2} (a) $|\hat{e}_{\rm \theta} \cdot \mathbf{\widetilde{E}}(\omega)|$ versus $\theta$, for $\nu = 100$ MHz, $E_{\rm C} = 10$ EeV, and $R = 2000$ m, accounting for LPM elongation, in the pure electromagnetic (solid black) and pure hadronic (solid gray) cases.  Shown for comparison are the pure electromagnetic (dashed black) and pure hadronic (dashed gray) from Fig. 1 of \cite{PhysRevD.61.023001}. (b) $|\hat{e}_{\rm \theta} \cdot \mathbf{\widetilde{E}}(\omega)|$ versus $\theta$, for $\nu = 100$ MHz, $E_{\rm C} = 1$ EeV, and $R = 500$ m, accounting for LPM elongation, in the pure electromagnetic (solid black) and pure hadronic (solid gray) cases.  Shown for comparison are the mixed (dashed black) and pure hadronic (dashed gray) from Fig. 4 of \cite{PhysRevD.61.023001}.  (c) Same as (a), but with $\nu = 1000$ MHz.  (d) Same as (b), but with $\nu = 1000$ MHz.  Notice that for $\nu = 1000$ MHz, the electromagnetic cone-width is larger than the hadronic cone-width.  See text for details.}
\end{figure}

Finally, $|\hat{e}_{\rm \theta} \cdot \mathbf{\widetilde{E}}|$ versus $\theta$ is compared with ZHS and AVZ \cite{PhysRevD.61.023001} in Fig. \ref{fig:lpm2}.  As with Fig. \ref{fig:lpm}, a counter-intuitive situation arises when $\eta > 1$.  Similar to the derivation of Eq. \ref{eq:width} from Eq. \ref{eq:money4}, the angular distribution at fixed frequency is Gaussian in $\Delta\cos\theta = \cos\theta-\cos\theta_{\rm C}$: $|\mathcal{W}(\eta,\theta)| \propto \exp\left(-\frac{1}{2}(\Delta\cos\theta/\sigma_{\rm \Delta\theta})^2\right)$, with Cherenkov cone-width

\begin{equation}
\sigma_{\rm \Delta\theta}(a,\nu,\eta) = \frac{c}{2\pi a\nu}(1+\eta^2)^{1/2}
\label{eq:width2}
\end{equation}

LPM elongation has been applied in Fig. \ref{fig:lpm2}, implying that the width of the Cherenkov cone-width should be narrower than ZHS curves.  In panel (a), the electromagnetic case agrees with AVZ at 100 MHz ($E_{\rm C} = 10$ EeV), but is narrower than ZHS, as expected.  The AVZ field for large $|\theta - \theta_{\rm C}|$ corresponds to radiation from sub-dominant peaks in the cascade profile not modelled by RB.  The LPM effect is not applied to the hadronic case in the associated code \cite{AlvarezMuniz1998396}.  The hadronic case is therefore wider in panel (a), because $\eta < 1$ and from Eq. \ref{eq:width2}, $\sigma_{\rm \Delta\theta} \sim 1/a\nu$.  This result follows past far-field approximations in the literature: the Cherenkov-cone is narrower for higher frequencies, and narrower for larger $a$.  However, pure hadronic cone-widths from the code should be narrower than AVZ pure hadronic cases, because AVZ do not have the $1/a$ dependence of Eq. \ref{eq:width2}.  In panel (b) ($E_{\rm C} = 1$ EeV), the code is compared to a mixed hadronic/electromagnetic cascade from \cite{PhysRevD.61.023001}.  AVZ note that pure electromagnetic cascades do not result from electroweak neutrino interactions, because the charged-current case always has a hadronic component.  Thus, a linear combination of the solid black and solid gray curves that sum to the black dashed curve will be observed for charged-current interactions.  The lower energy in panel (b) relative to panel (a) leads to a wider cone-width because LPM elongation is proportional to $E_{\rm C}$.

In panel (c) of Fig. \ref{fig:lpm2}, the same scenario as in panel (a) is depicted, but the frequency is an order of magnitude larger.  In panel (d), the same scenario as in panel (b) is depicted, but the frequency is an order of magnitude larger as well.  For both scenarios, $\eta \gg 1$, so the Cherenkov cone \textit{loses dependence on frequency:} $\sigma_{\rm \Delta\theta} \sim \frac{a}{R}\sin^2\theta$.  This result is that the electromagnetic contribution to the Cherenkov cone is actually \textit{wider} than the hadronic contribution, because the LPM elongation increases the $a$ of the electromagnetic component, but not the hadronic component.  Note that the solid black and solid gray curves do not have to agree with the dashed curves, because they represent different frequencies.  The AVZ and ZHS results are identical in all panels, and kept for reference in panels (c) and (d).

\section{Time-Domain Properties at the Cherenkov Angle}
\label{sec:time}

The analytic RB+LPM+$\widetilde{F}(\omega,\theta)$ model is derived in the time-domain for limiting cases, and parameters from the semi-analytic treatment in Ref. \cite{PhysRevD.84.103003} are derived analytically.  The purpose of this section is to connect each parameter in the analytic Askaryan pulse to a physical origin in the cascade.  Similar efforts have been attempted.  The authors of \cite{Razz} provide a formula along the lines of Eqs. \ref{eq:form7} and \ref{eq:form12} involving Bessel function evaluation, but rely on MC techniques to complete the model.  The authors of \cite{YuHu} chose a mathematically tractable ICD and connected ICD parameters to simulated Askaryan radiation properties.  However, the choice of ICD in \cite{YuHu} was for convenience, and does not necessarily agree with MC results.  Here, we have attempted to both choose a 3D ICD that matches Geant4 results, and evaluate the Askaryan field analytically.  See Appendix (Sec. \ref{sec:form1}) for further detail.

Two cases are considered: $\widetilde{F} = 1$, followed by $\widetilde{F}(\omega,\theta) \neq 1$.  The limiting frequency of the former, $\omega_{\rm C}$, is governed by coherence.  The latter has two limiting frequencies, $\omega_{\rm C}$ and $\omega_{\rm CF}$, which leads to an asymmetry in the vector potential, and therefore, asymmetry in $\widetilde{\textbf{E}}$.  The SI units of terms like $R\mathbf{\widetilde{E}}$ in the Fourier domain are $\left[\SI{}{\volt}/\SI{}{\hertz}\right]$, while they are just $\left[\SI{}{\volt}\right]$ for $R\mathbf{E}$ in the time-domain.  The overall scale of the field is not relevant in this section, so the unit of frequency is left as $\left[\SI{}{\hertz}\right]$, rather than $\left[\SI{}{\MHz}\right]$.  In each derivation, the viewing angle is $\theta = \theta_{\rm C}$.

\subsection{The limit $\eta < 1$, $\widetilde{F}(\omega,\theta_{\rm C})$ = 1}
\label{sec:time1}

Recall from Eq. \ref{eq:money} of Sec. \ref{sec:rb0} that the vector-form of the on-cone field from the RB formalism takes the form:

\begin{equation}
\frac{R\mathbf{\widetilde{E}}(\omega,\theta_{\rm C})}{\left[\SI{}{\volt}/\SI{}{\hertz}\right]} = - \frac{i \omega E_{\rm 0} \sin\theta_{\rm C} e^{i\omega R/c}}{(1-i\eta)^{1/2}} \hat{e_{\rm \theta}}
\label{eq:eq0}
\end{equation}

Let $\hat{E_{\rm 0}} = E_{\rm 0} \sin\theta_{\rm C} \hat{e}_{\theta}$, and define $\omega_C$ from $\eta$: $\eta = \omega/\omega_{\rm C}$.  Equation \ref{eq:eq0} may be approximated to first order in the limit $\eta < 1$, or $\omega < \omega_{\rm C}$, equivalent to requiring $\lambda R \gtrsim 5a^2$.  Using the definition of $\eta$, $\nu_{\rm C} = \omega_{\rm C}/(2\pi)$ is

\begin{equation}
\label{eq:eqA}
\nu_{\rm C} = \frac{c R}{2\pi a^2 \sin^2\theta_{\rm C}}
\end{equation}

Applying the given limit to Eq. \ref{eq:eq0}, and taking the inverse Fourier transform, yields

\begin{equation}
\label{eq:eq1}
R\mathbf{E}(t_{\rm r},\theta_{\rm C}) \approx \frac{i\omega_{\rm C}\hat{E_{\rm 0}}}{\pi}\frac{d}{dt_{\rm r}} \int_{-\infty}^{\infty} d\omega \frac{e^{-it_{\rm r}\omega}}{\omega + 2 i \omega_{\rm C}}
\end{equation}

The sign convention in the exponential in Eq. \ref{eq:eq1} is chosen to remain consistent with the RB formalism.  The integral may be performed using the Cauchy integral formula, provided that the numerator is analytic ($\exp(-i \omega t_{\rm r})$ obeys the Cauchy-Riemann equations).

Contour integration of Eq. \ref{eq:eq1} requires a contour $C$ that satisfies Jordan's lemma and includes all real omega: $\omega \in [-\infty,\infty]$.  For the $t_{\rm r} > 0$ case, the integral converges along the contour defined by the infinite lower semi-circle because the magnitude of the numerator decreases like $\exp(\operatorname{Im}\lbrace \omega \rbrace)$.  Note that this is a negatively-oriented contour.  For the case $t_{\rm r} < 0$, use the fact that $\mathcal{F}_\omega(x(-t)) = \widetilde{X}(-\omega)$, so $x(-t) = \mathcal{F}_\omega^{-1}(\widetilde{X}(-\omega))$, where $\mathcal{F}_\omega(x) = {\widetilde{X}(\omega)}$ is the Fourier transform of a function $x(t)$.  The final solution is piecewise:

\begin{equation}
\frac{R \textbf{E}(t_{\rm r},\theta_{\rm C})}{\left[\SI{}{\volt}\right]} \approx 4 \hat{E_{\rm 0}} \omega_C^2
\begin{cases} 
      \exp(2\omega_{\rm C} t_{\rm r}) & t_{\rm r} \leq 0 \\
      -\exp(-2\omega_{\rm C} t_{\rm r}) & t_{\rm r} > 0
\end{cases}
\label{eq:eq3}
\end{equation}

\begin{figure}
\centering
\includegraphics[width=0.5\textwidth,trim=0cm 5.5cm 0cm 6cm,clip=true]{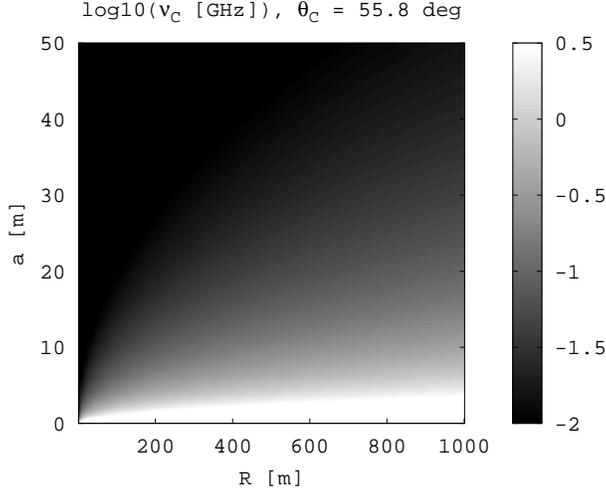}
\caption{\label{fig:cutoff1} A contour graph of $\log_{\rm 10}\nu_{\rm C}$, for a parameter space relevant for ground-based radio-Askaryan detectors.  Notice that $\nu_{\rm C}<1$ GHz if $a\gtrsim 1$ m.}
\end{figure}

MC calculations show the transition at $t_{\rm r} = 0$ to be smooth \cite{Eug}.  Equation \ref{eq:eq3} has a characteristic width of $1/\omega_{\rm C} = 1/(2\pi\nu_{\rm C})$, \textit{implying that the pulse-width is controlled by coherence, in the absence of a form factor}.  Figure \ref{fig:cutoff1} shows $\nu_{\rm C}$ versus the observer distance $R$ and the shower width $a$.

Under the Lorentz gauge condition for Maxwell's equations, in the absence of static potentials, the negative derivative of the vector potential yields the electric field: $-\partial \textbf{A}/\partial t = \textbf{E}$.  Using Eq. \ref{eq:eq3}, the vector potential is

\begin{equation}
\frac{R \textbf{A}(t_{\rm r},\theta_{\rm C})}{\left[\SI{}{\volt}\cdot\SI{}{\second}\right]} \approx -2 \hat{E_{\rm 0}} \omega_{\rm C} \\
\begin{cases} 
      \exp(2\omega_{\rm C} t_{\rm r}) & t_{\rm r} \leq 0 \\
      \exp(-2\omega_{\rm C} t_{\rm r}) & t_{\rm r} > 0
\end{cases}
\label{eq:eq4}
\end{equation}

Equation 16 of \cite{PhysRevD.84.103003} is the vector potential at $\theta = \theta_{\rm C}$: 

\begin{equation}
\frac{R \textbf{A}(t_{\rm r},\theta_{\rm C})}{\left[\SI{}{\volt}\cdot\SI{}{\second}\right]} = - E_{\rm 0}' \sin(\theta_{\rm C}) \hat{e_{\rm \theta}} \left(\exp(-2|t_{\rm r}|/x_{\rm 0})+(1+x_{\rm 1}|t_{\rm r}|)^{-x_{\rm 2}} \right)
\label{eq:eq5}
\end{equation}

Eq. \ref{eq:eq5} is a formula that is used in MC by ARA/ARIANNA \cite{Allison2012457} \cite{Eug} \cite{HansonThesis}, corresponding to a fit to MC similar to ZHS.  By comparing Eqs. \ref{eq:eq4} and \ref{eq:eq5}, a natural, theoretical explanation of the fit parameters in \cite{PhysRevD.84.103003} arises, albeit from a special case: $\widetilde{F} = 1$, and $x_{\rm 0} = 1/(\omega_{\rm C})$, with $x_{\rm 2} \gg 1$, or $x_{\rm 1} \sim 0$.  Thus, the result from \cite{PhysRevD.84.103003} has been derived from first principles, rather than fitting to MC.

The fits in Ref. \cite{PhysRevD.84.103003} have $x_{\rm 2} \approx x_{\rm 1} \approx 3$.  The fact that $x_{\rm 1}$ and $x_{\rm 2}$ are not relevant to Eq. \ref{eq:eq4} is precisely because stipulating that $\widetilde{F}(\omega,\theta_{\rm C})=1$ leaves the spectral limiting to $\nu_{\rm C}$ rather than $\widetilde{F}(\omega,\theta)$.  Such a scenario can be important when dealing with observations of cascades with $R = \mathcal{O}(100)$ m, under the influence of the LPM effect.  In this case, only a small fraction of the shower excess profile is within $\Delta z_{\rm coh}$, and $\nu_{\rm C}$ cuts off the spectrum.  Another example in which spectral limiting is due to $\nu_{\rm C}$, rather than $\widetilde{F}(\omega,\theta)$, is when the dielectric medium is denser than ice.  The Askaryan spectra extends to $\approx\mathcal{O}(10)$ GHz at $\theta=\theta_{\rm C}$ in salt, for example \cite{PhysRevD.74.043002}.  Higher density leads to a more compact ICD, suppressing the effect of $\widetilde{F}(\omega,\theta)$.  Figure \ref{fig:cutoff1} shows a parameter space for $\nu_{\rm C}$ relevant to ARA/ARIANNA.

The result $x_{\rm 0} = 1/(\omega_{\rm C})$ also has a useful physical analogy for the shower width, $a$.  Let the signal propagation time be $T$, such that (to first order) $R = cT/n$.  Equation \ref{eq:eq6} then relates the pulse width $x_{\rm 0}$ from Eq. \ref{eq:eq5} to the shower width $a$:

\begin{equation}
\label{eq:eq6}
x_{\rm 0} = \left( \frac{a\sin\theta_{\rm C}}{c} \right) \left( \frac{a\sin\theta_{\rm C}}{R} \right) = T\left(\frac{a\sin\theta_{\rm C}}{\sqrt{n}R}\right)^2
\end{equation}

Equation \ref{eq:eq6} demonstrates that the pulse width is a fraction of the propagation time $T$, and proportional to $(a/R)^2$.

\subsection{The limit $\eta < 1$, $\sigma < 1$, $\widetilde{F}(\omega,\theta_{\rm C}) \neq 1$}
\label{sec:time2}

Askaryan radiation from cascades experiences further low-pass filtering from $\widetilde{F}(\omega,\theta) \neq 1$ (Sec. \ref{sec:form}).  The parameter $\sigma$ can be used to define a limiting frequency: $\sigma = \omega/\omega_{\rm CF}$, similar to $\eta = \omega/\omega_{\rm C}$.  The electric field of Eq. \ref{eq:eq0}, combined with the form factor $\widetilde{F}(\omega,\theta_{\rm C})$ of Eq. \ref{eq:form7}, is

\begin{equation}
\label{eq:eq7}
\frac{R \mathbf{\widetilde{E}}(\omega,\theta_{\rm C})}{\left[\SI{}{\volt}/\SI{}{\hertz}\right]} = - \widetilde{F}(\omega,\theta) \frac{i \omega E_{\rm 0} \sin\theta_{\rm C} e^{i\omega R/c}}{(1-i\omega/\omega_{\rm C})^{1/2}} \hat{e_{\rm \theta}} = - \frac{i \omega E_{\rm 0} \sin\theta_{\rm C} e^{i\omega R/c}}{(1-i\omega/\omega_{\rm C})^{1/2}(1+(\omega/\omega_{\rm CF})^2)^{3/2}} \hat{e_{\rm \theta}}
\end{equation}

In the limit $\sigma < 1$, and $\eta < 1$, Eq. \ref{eq:eq7} may be approximated as shown in Eq. \ref{eq:eq8}, using $t_{\rm r} = t - R/c$, and $\omega_{\rm 0} = \sqrt{2/3}~\omega_{\rm CF}$.

\begin{equation}
\label{eq:eq8}
\frac{R \mathbf{E}(t_{\rm r},\theta_{\rm C})}{\left[\SI{}{\volt}\right]} \approx \frac{i \omega_{\rm 0}^2 \omega_{\rm C}}{\pi} \hat{E_{\rm 0}} \frac{d}{dt_{\rm r}} \oint d\omega \frac{e^{-it_{\rm r}\omega}}{(\omega+2i\omega_{\rm C})(\omega+i\omega_{\rm 0})(\omega-i\omega_{\rm 0})}
\end{equation}

There are two poles in the lower-half complex plane, and one in the upper-half plane.  If $t_{\rm r}>0$, the contour integral around the lower infinite semi-circle converges because the numerator approaches zero exponentially as $\operatorname{Im}\lbrace{\omega}\rbrace \rightarrow -\infty$.  Conversely for $t_{\rm r}<0$, the contour integral converges along the upper infinite semi-circle.  The final field is given by Eq. \ref{eq:eq9}, to first-order in $\epsilon$, with $\epsilon = \omega_{\rm 0}/\omega_{\rm C}$.

\begin{equation}
\label{eq:eq9}
\frac{R \mathbf{E}(t_{\rm r},\theta_{\rm C})}{\left[\SI{}{\volt}\right]} \approx \frac{\hat{E_{\rm 0}}\omega_{\rm CF}^2}{3}
\begin{cases} 
      (1-\frac{1}{2}\epsilon)\exp(\omega_{\rm 0} t_{\rm r}) & t_{\rm r} \leq 0 \\
      -\exp(-\omega_{\rm 0} t_{\rm r})+2\exp(-2\omega_{\rm C} t_{\rm r}) & t_{\rm r} > 0
\end{cases}
\end{equation}

\begin{figure}
\centering
\includegraphics[width=0.5\textwidth,trim=0cm 5.5cm 0cm 6cm,clip=true]{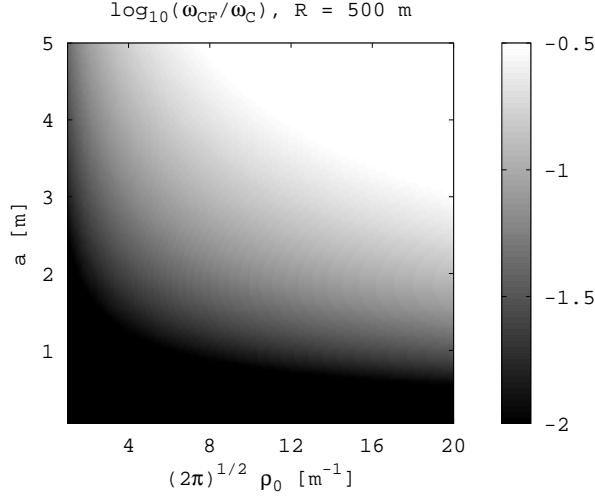}
\caption{\label{fig:cutoff2} A contour plot of $\omega_{\rm CF}/\omega_{\rm C} = \nu_{\rm CF}/\nu_{\rm C}$, for a parameter space relevant for ground-based radio-Askaryan detectors.}
\end{figure}

Consulting Fig. \ref{fig:cutoff1} reveals regions of parameter space where $\omega_{\rm C} \leq 1$ GHz.  Consulting Eq. \ref{eq:form4} and Eq. \ref{eq:eqA} shows that $\epsilon < 1$ is typical for cascades with $a\leq\mathcal{O}(1-10)$ m.  The relative strengths of $\omega_{\rm C}$ and $\omega_{\rm CF} = \sqrt{3/2}\omega_{\rm 0}$ are shown in Fig. \ref{fig:cutoff2}, versus the longitudinal and lateral cascade widths.  The vector potential corresponding to Eq. \ref{eq:eq9} is

\begin{equation}
\label{eq:eq11}
\frac{R \textbf{A}(t_{\rm r},\theta_{\rm C})}{\left[\SI{}{\volt}\cdot\SI{}{\second}\right]} \approx -\frac{\hat{E_{\rm 0}}\omega_{\rm CF}}{\sqrt{6}}
\begin{cases} 
	(1-\frac{1}{2}\epsilon)\exp(\omega_{\rm 0} t_{\rm r}) & t_{\rm r} \leq 0 \\
	\exp(-\omega_{\rm 0} t_{\rm r}) - \epsilon\exp(-2\omega_{\rm C} t_{\rm r}) & t_{\rm r} > 0
\end{cases}
\end{equation}

Equations \ref{eq:eq9} and \ref{eq:eq11} show that the field remains bipolar but asymmetric, and asymmetric in time, from the interplay between coherence and the form factor. The pulse width is enhanced due to the presence of two different limiting frequencies, $\omega_{\rm 0} = \sqrt{2/3} \omega_{\rm CF}$, and $\omega_{\rm C}$.  Equation \ref{eq:eq12} defines a parameter showing the relative importance of the two limiting frequencies:

\begin{equation}
\label{eq:eq12}
\epsilon' = \omega_{\rm CF}/\omega_{\rm C} = (\sqrt{2\pi}\rho_{\rm 0} \rho) \left(\frac{a}{R}\right)^2
\end{equation}

Fig. \ref{fig:cutoff2} is a contour graph of $\epsilon'$ in a parameter space relevant for ARA/ARIANNA.  The first term in parentheses in Eq. \ref{eq:eq12} represents the relative importance of $\widetilde{F}(\omega,\theta)$.  The second term in parentheses is the ratio of the longitudinal cascade width to the observer distance, and it represents the quality of the Fraunhofer limit.

\subsection{Generalization of Eq. \ref{eq:eq7}}
\label{sec:time3}

The purpose of Eq. \ref{eq:form12} is to account for excess charge at lateral distances greater than one Moli\`{e}re radius.  Consequently, Eq. \ref{eq:eq7} may be generalized to

\begin{equation}
\label{eq:13}
\frac{R \mathbf{\widetilde{E}}(\omega,\theta_{\rm C})}{\left[\SI{}{\volt}/\SI{}{\hertz}\right]} = - \sum_{j=0}^N \frac{\alpha_{\rm j}i \omega \hat{E}_{\rm 0} e^{i\omega R/c}}{(1-i\omega/\omega_{\rm C})^{1/2}(1+(\omega/\omega_{\rm CF,j})^2)^{3/2}}
\end{equation}

Making the same approximations leading up to Eq. \ref{eq:eq8}, and exchanging the order of summation and integration, yields

\begin{equation}
\label{eq:eq14}
\frac{R \mathbf{E}(t_{\rm r},\theta_{\rm C})}{\left[\SI{}{\volt}\right]} \approx \sum_{\rm j=0}^N \alpha_{\rm j} \frac{i \omega_{\rm C} \omega_{\rm 0,j}^2}{\pi} \hat{E_{\rm 0}} \frac{d}{dt_{\rm r}} \oint d\omega \frac{e^{-it_{\rm r}\omega}}{(\omega+2i\omega_{\rm C})(\omega+i\omega_{\rm 0,j})(\omega-i\omega_{\rm 0,j})}
\end{equation}

Thus, accounting for wider lateral scales in the ICD (here in the far-field case) amounts to adding more poles in the complex-$\omega$ plane.  The problem mirrors the addition of complex poles to digital filters to achieve a desired filter impulse response.  The summation terms of Eq. \ref{eq:eq14} are equal to Eq. \ref{eq:eq9}, with the substitution $\omega_{\rm 0} \rightarrow \omega_{\rm 0,j}$:

\begin{equation}
\label{eq:eq15}
\frac{R \mathbf{E}(t_{\rm r},\theta_{\rm C})}{\left[\SI{}{\volt}\right]} \approx \sum_{\rm j=0}^N \alpha_{\rm j} \frac{\hat{E_{\rm 0}}\omega_{\rm 0,j}^2}{2}
\begin{cases} 
      (1-\frac{1}{2}\frac{\omega_{\rm 0,j}}{\omega_{\rm C}})\exp(\omega_{\rm 0,j} t_{\rm r}) & t_{\rm r} \leq 0 \\
      -\exp(-\omega_{\rm 0,j} t_{\rm r})+2\exp(-2\omega_{\rm C} t_{\rm r}) & t_{\rm r} > 0
\end{cases}
\end{equation}

The pair of poles in Eq. \ref{eq:eq7} from the form factor $\widetilde{F}$ follow from fitting the $\rho'$-component of the ICD with a single exponential function.  The next-most complex case is to fit the ICD $\rho'$-component with two exponentials ($N = 1$ in Eq. \ref{eq:eq15}, with $\Delta_{\rm 0,j} = 2\omega_{\rm C} - \omega_{\rm 0,j}$):

\begin{equation}
\label{eq:eq16}
\frac{R \mathbf{E}(t_{\rm r},\theta_{\rm C})}{\left[\SI{}{\volt}\right]} \approx
\begin{cases} 
      \hat{E}_{\rm 0}\left( \alpha_{\rm 0}\omega_{\rm 0,0}\epsilon_{\rm 0}\Delta_{\rm 0,0}e^{\omega_{\rm 0,0}t_{\rm r}} + \alpha_{\rm 1}\omega_{\rm 0,1}\epsilon_{\rm 1}\Delta_{\rm 0,0}e^{\omega_{\rm 0,1}t_{\rm r}} \right) & t_{\rm r} \leq 0 \\
      \hat{E}_{\rm 0}e^{-2\omega_{\rm C}t_{\rm r}}\left( \alpha_{\rm 0}\omega_{\rm 0,0}^2\left(1-\frac{1}{2}e^{-\Delta_{\rm 0,0}t_{\rm r}}\right) + \alpha_{\rm 1}\omega_{\rm 0,1}^2\left(1-\frac{1}{2}e^{-\Delta_{\rm 0,1}t_{\rm r}}\right) \right) & t_{\rm r} > 0
\end{cases}
\end{equation}

\section{Summary}

The Askaryan fields for a UHE-$\nu$ induced cascade have been presented, accounting for LPM elongation, and the 3D ICD of the cascade.  The calculations are analytic, and the associated code encapsulates them into a model that is computationally efficient and in agreement with prior studies.  The entire field in both $\hat{e}_{\rm r}$ and $\hat{e}_{\rm \theta}$ components is computed in the code, for any set of initial conditions chosen by users.  The cascade model was verified with Geant4, and the 3D MC cascade structure formed the shape of the 3D ICD model.  Evaluating the spatial Fourier transform of the 3D ICD yielded the form factor, $\widetilde{F}(\omega,\theta)$.  The effects of $\widetilde{F}(\omega,\theta)$ and LPM elongation were explored mathematically and graphically.  Table \ref{tab:summary} contains brief summary of the results and tools presented in this work.

\begin{table}
\begin{center}
\begin{tabular}{c c c}
\hline
Effect & Eq./Fig. & Sec. \\ \hline
Cascade Form Factor, $\widetilde{F}(\omega,\theta)$, single-pole & Eq. \ref{eq:form7} & Sec. \ref{sec:form2} \\
Cascade Form Factor, $\widetilde{F}(\omega,\theta)$, multi-pole & Eq. \ref{eq:form12} & Sec. \ref{sec:form3} \\
Contours of complete model (RB+LPM+$\widetilde{F}(\omega,\theta)$) $\hat{e}_{\theta} \cdot \widetilde{\mathbf{E}}(t,\theta)$ & Fig. \ref{fig:contours1} & Sec. \ref{sec:contours} \\ 
Contours of complete model (RB+LPM+$\widetilde{F}(\omega,\theta)$) $\hat{e}_{r} \cdot \widetilde{\mathbf{E}}(t,\theta)$ & Fig. \ref{fig:contours3} & Sec. \ref{sec:contours} \\ 
Complete model (RB+LPM+$\widetilde{F}(\omega,\theta)$) $\hat{e}_{\theta} \cdot \widetilde{\mathbf{E}}(\omega)$ comparison & Fig. \ref{fig:lpm} & Sec. \ref{sec:rb1} \\
Complete model (RB+LPM+$\widetilde{F}(\omega,\theta)$) $\hat{e}_{\theta} \cdot \widetilde{\mathbf{E}}(\theta)$ comparison & Fig. \ref{fig:lpm2} & Sec. \ref{sec:rb1} \\ 
$\mathbf{E}(t,\theta_{\rm C})$, $\widetilde{F}(\omega,\theta)=1$, $\eta < 1$ & Eq. \ref{eq:eq3} & Sec. \ref{sec:time1} \\
$\mathbf{E}(t,\theta_{\rm C})$, $\widetilde{F}(\omega,\theta)\neq 1$, $\eta < 1$, single-pole & Eq. \ref{eq:eq9} & Sec. \ref{sec:time2} \\
$\omega_{\rm CF}/\omega_{\rm C}$ Figure of Merit & Eq. \ref{eq:eq12} & Sec. \ref{sec:time2} \\
$\mathbf{E}(t,\theta_{\rm C})$, $\widetilde{F}(\omega,\theta)\neq 1$, $\eta < 1$, two-pole & Eq. \ref{eq:eq16} & Sec. \ref{sec:time3} \\
\hline
\end{tabular}
\end{center}
\caption{\label{tab:summary} A summary of new results.}
\end{table}

The effect of LPM elongation was modelled as an energy-dependent increase in the longitudinal cascade width, $a$.  LPM elongation is found to modify low-frequency emission, to suppress high-frequency emission, and to narrow the Cherenkov cone under far-field conditions.  $\widetilde{F}(\omega)$ is similar to a two-pole, low-pass filter, with the limiting frequency determined by cascade Moli\`{e}re radius and viewing angle $\theta$.  The $\theta$-dependence in the form of $\widetilde{F}(\omega,\theta)$ implies that the filtered radiation depends on the laterally-projected wavevector.  Although these conclusions are in line with expectations, alternate scenarios were explored, yielding novel results (see Figs. \ref{fig:lpm} and \ref{fig:lpm2}).  The field shows interesting causal structure that could serve as a discrimination technique between the distance $R$ and the cascade energy, for \textit{in situ} detectors such as ARA/ARIANNA (see Figs. \ref{fig:contours1} and \ref{fig:contours3}).

Finally, time-domain field equations were derived by computing the inverse Fourier transform of the RB model.  Future work will focus exclusively on the time domain, for viewing angles $\theta \neq \theta_{\rm C}$, and expanded ranges of $\eta$ and $\omega$.  Producing theoretical, time-dependent field equations under specific limits facilitates UHE-$\nu$ signal template generation by bypassing altogether the need for Askaryan RF code, provided the limits are satisfied.  Rejecting thermal noise in the \textit{in situ} Antarctic detectors, in favour of UHE-$\nu$ signals, is an exercise in the mathematical analysis of thermal fluctuations \cite{6771565}.  The associated code presented in this work has been made freely available \footnote{https://github.com/918particle/AskaryanModule} to all researchers involved in discriminating cosmogenic UHE-$\nu$ pulses from thermal backgrounds.

\section{Acknowledgements}

We thank the NSF for the support received under the NSF CAREER award 1255557, and for support from the NSF award for the Askaryan Radio Array (Grant 1404266).  We are grateful to the Ohio Supercomputing Center \cite{OhioSupercomputerCenter1987} for their staff support, training, and resources.  Contributions were also received from The Ohio State University, the Center for Cosmology and Astroparticle Physics (CCAPP), and the United States-Israel Bilateral Science Foundation (BSF) Grant 2012077.  We thank John Ralston and Roman Buniy for their helpful remarks regarding the mathematical physics.  Also, special thanks are in order for John Beacom and Steve Barwick for practical advice.

\section{Appendix}

\subsection{Causal Features and Poles of Askaryan Radiation}
\label{sec:caus}

The complex pole-structure of the various models each demonstrate how the models treat the issue of causality.  The E-field of ZHS, on-cone in Eq. \ref{eq:zhs1} takes the form

\begin{equation}
\label{eq:zhs3}
\frac{R\mathbf{\widetilde{E}}}{\left[\SI{}{\volt}/\SI{}{\Hz}\right]} = -E_{\rm 0}\omega_{\rm 0}^2\frac{i\omega}{(\omega+i\omega_{\rm 0})(\omega-i\omega_{\rm 0})} \hat{e}_{\rm \theta}
\end{equation}

Figure 16 of ZHS shows that the E-field phase is $\sim 90^{\circ}$ below 1 GHz, or a phase factor of $\exp(i\pi/2) = i$.  The overall minus sign in Eq. \ref{eq:zhs3} is just a convention.  Taking the inverse Fourier transform, the time-domain form of the field at the Cherenkov angle may be written

\begin{equation}
\label{eq:zhs4}
\frac{R\mathbf{E}(t)}{\left[\SI{}{\volt}\right]} = \omega_{\rm 0}^2 E_{\rm 0} \frac{d}{dt}\int_{-\infty}^{\infty} \frac{e^{-i\omega t}}{(\omega+i\omega_{\rm 0})(\omega-i\omega_{\rm 0})} d\omega
\end{equation}

The integral converges via Jordan's lemma if the contour is the infinite upper semi-circle for $t<0$, and, for $t>0$, the infinite lower semi-circle.  There is an overall minus sign from the clockwise contour.  The result is

\begin{equation}
\label{eq:zhs5}
\frac{R\textbf{E}(t)}{ \left[\SI{}{\volt}\right]} = \omega_{\rm 0}^2 E_{\rm 0} \hat{e_{\rm \theta}}
\begin{cases} 
      \exp(\omega_{\rm 0} t) & t \leq 0 \\
      -\exp(-\omega_{\rm 0} t) & t > 0
\end{cases}
\end{equation}

The existence poles above and below the real line is deemed a causality violation by RB.  Physically, the field changes overall sign when the angular acceleration of the charge relative to the observer changes sign.  Feynman's formula \cite{PhysRevD.45.362} states that the field from an accelerating charge goes like $\mathbf{E} \propto \sgn(1-n\hat{u}\cdot\vec{\beta}) \hat{u} \times \ddot{\theta}$, where $\vec{\beta}$ is the velocity of the charge, and $\hat{u}$ is a unit vector at the charge location in the direction of the observer.  $\mathbf{E}$ changes sign as the charge crosses the plane in which $R$ is minimized.  The quantity $\ddot{\theta}$ increases rapidly, until the plane crossing, after which it decreases rapidly.

From Eq. \ref{eq:zhs4} $\widetilde{F}_{\rm ZHS}(\omega) \propto (\omega +i\omega_{\rm 0})^{-1}(\omega - i\omega_{\rm 0})^{-1}$.  Treating $t>0$ and $t<0$ separately, the inverse Fourier transform of $\widetilde{F}_{\rm ZHS}(\omega)$ with respect to the coordinate $\rho'$ yields $f(\mathbf{x}') \propto \exp(-\rho')$.  Therefore, a logical inference is that \textit{the full, 3D ICD responsible for $\widetilde{F}(\omega,\theta)$  is distributed exponentially}.  Geant4 simulations show this to be correct in Sec. \ref{sec:form1}.

\subsection{Numerical Study of the Excess Charge Distribution}
\label{sec:form1}

Geant4 \cite{geant1} \cite{geant2} is used to derive numbers for $\sqrt{2\pi}\rho_{\rm 0}$, and those results are checked with Eqs. \ref{eq:formG1}-\ref{eq:formG3}.  Refs. \cite{Razz} \cite{Alv2003} \cite{McKayHussain} are other works that used GEANT/Pythia to calculate Askaryan radiation properties.  The GEANT4 high-energy electromagnetic option-1 physics list was used, with a MC threshold of 1 MeV, $e^{\pm}$ primaries, and ice of density 0.917 g/cm$^{3}$ and at a temperature of 240 K.  Although the LPM effect is important primarily for electromagnetic cascades, $\widetilde{F}(\omega,\theta)$ does not depend on $a$, so it is also valid for hadronic cascades.

CPU memory constraints forbid accounting for all tracks, so a pre-shower/sub-shower approach is taken to access more memory.  A pre-shower drops all particles with energy below 0.1 PeV.  The trajectory, position and type of the pre-shower particles generated by the primary are recorded and sent to separate CPUs.  Each particle in the pre-shower then becomes an independent cascade, with a second MC threshold of 1 MeV.

\begin{figure}
\centering
\begin{subfigure}{0.48\textwidth}
\centering
\includegraphics[width=\textwidth,trim=1cm 2cm 1cm 3cm,clip=true]{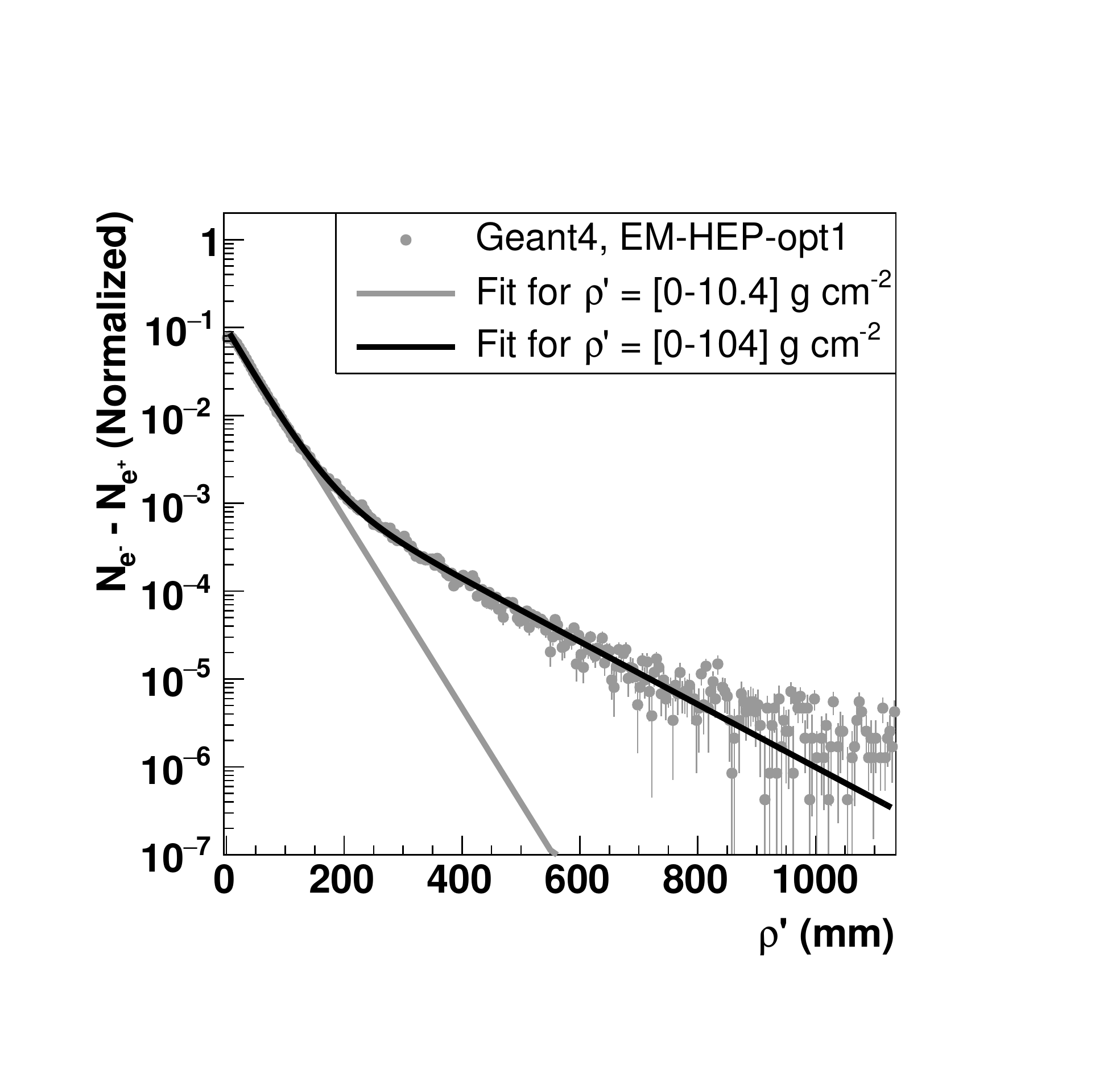}
\caption{}
\end{subfigure}%
\begin{subfigure}{0.48\textwidth}
\centering
\includegraphics[width=\textwidth,trim=0cm 6cm 1cm 6cm,clip=true]{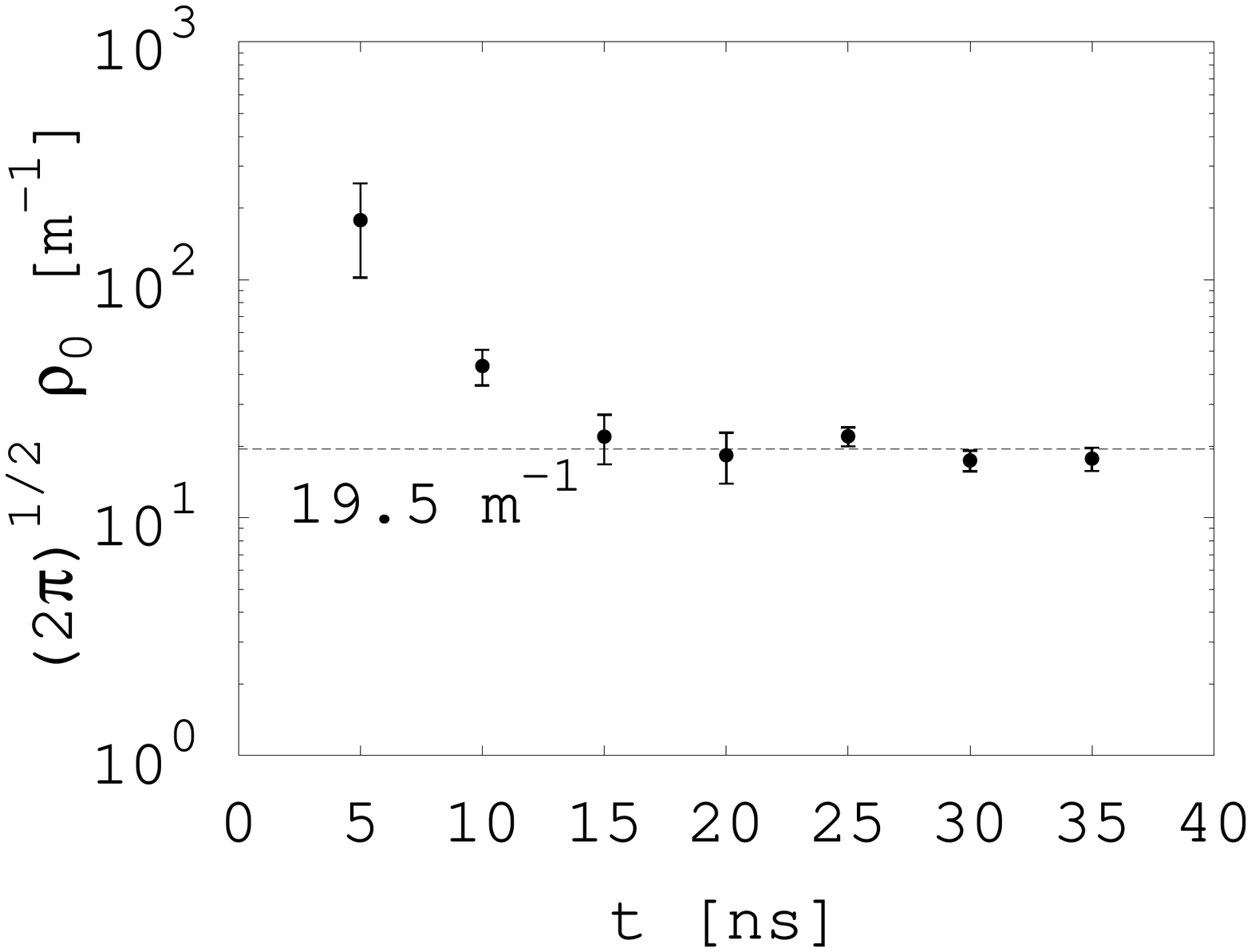}
\caption{}
\end{subfigure}
\caption{\label{fig:formG1} (a) $N_{\rm e^{-}} - N_{\rm e^{+}}$, versus $\rho'$ at 25 ns into a 100 PeV shower (gray points).  A single-exponential model (gray line, Eq. \ref{eq:form2}) with slope $\sqrt{2\pi}\rho_{\rm 0}$ is fit for $\rho' < 10.4$ g cm$^{-2}$.  A double-exponential model (black line, Eq. \ref{eq:form8}) is fit for $\rho' < 104$ g cm$^{-2}$. (b) Fit results for the parameter $\sqrt{2\pi}\rho_{\rm 0}$ vs. time within the shower, using Eq. \ref{eq:form2}.  The dashed line is the average of the points between 15-35 ns.}
\end{figure}

The lateral ICD is shown in Fig. \ref{fig:formG1} (a).  The results follow $\propto \exp(-\sqrt{2\pi}\rho_{\rm 0} \rho')$ in the range $\rho' = [0,\rho_{\rm 1}/d_{\rm ice}]$ (0-113 mm in ice, with $\rho_{\rm 1} = 10.4$ g cm$^{-2}$ and $d_{\rm ice} = 0.917$ g cm$^{-3}$).  The gray data corresponds to Geant4 tracks inside a 100~PeV cascade, $25\pm0.01$ ns from the beginning of the first Geant4 interaction.  The single exponential (single-pole) fit diverges when $\rho' > \rho_{\rm 1}$, however a double exponential (two-pole) model, comprised of a sum of exponential functions, fits the data for $\rho' > \rho_{\rm 1}$.

The results for $\sqrt{2\pi}\rho_{\rm 0}$ are shown in Fig. \ref{fig:formG1} (b), averaged over 10 cascades with $E_{\rm C} = 100$ PeV.  Each point contains tracks existing within 10 ps of the time on the x-axis.  Early in the cascade, the particles have not yet diffused laterally, implying a higher value of $\sqrt{2\pi}\rho_{\rm 0}$.  The dashed horizontal line represents the average between 15-35 ns, when lateral diffusion saturates.

\begin{figure}
\centering
\begin{subfigure}{0.48\textwidth}
\centering
\includegraphics[width=\textwidth,trim=1cm 1.5cm 1cm 3cm,clip=true]{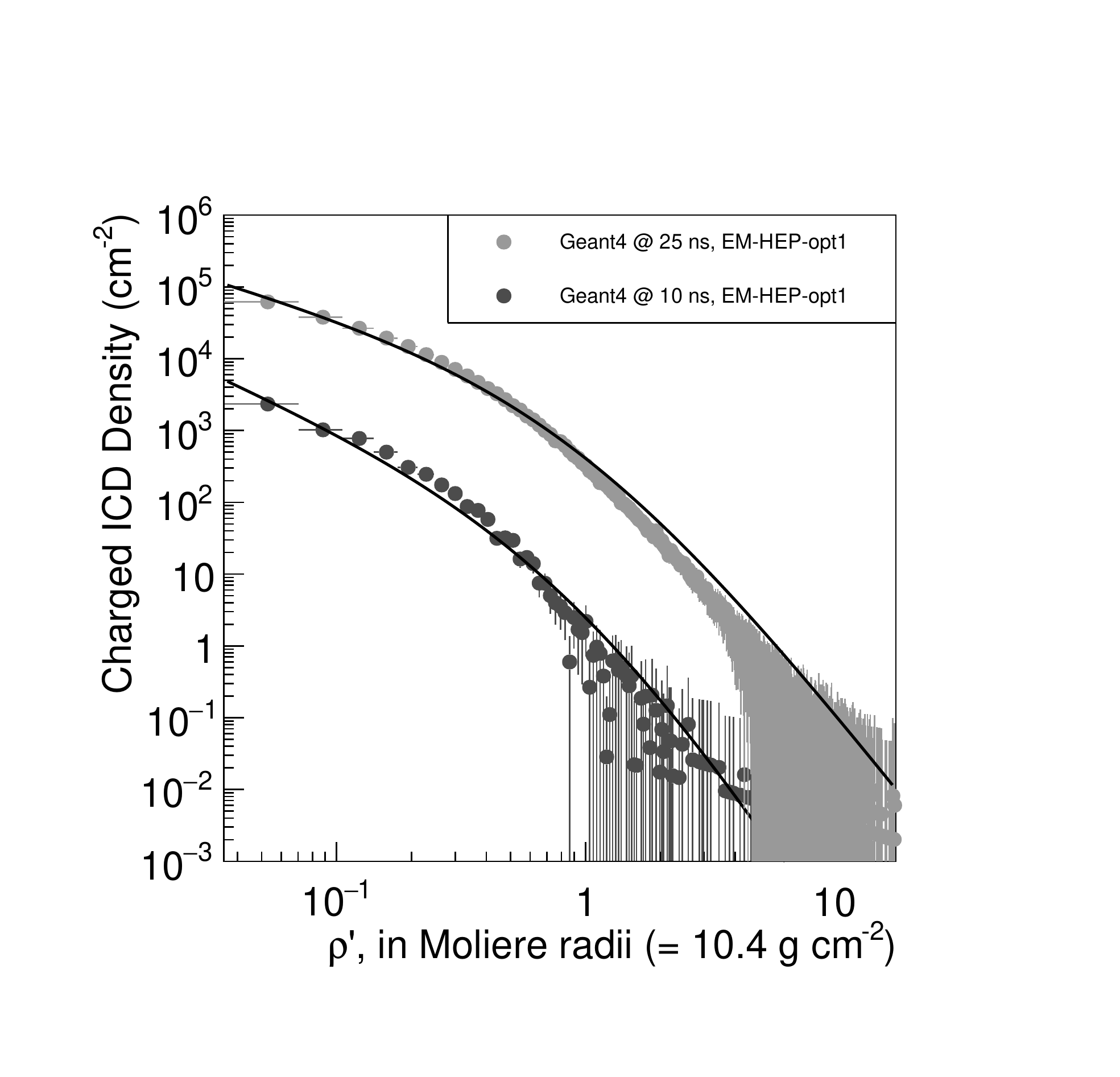}
\caption{}
\end{subfigure}%
\begin{subfigure}{0.48\textwidth}
\centering
\includegraphics[width=\textwidth,trim=1cm 6cm 1cm 6cm,clip=true]{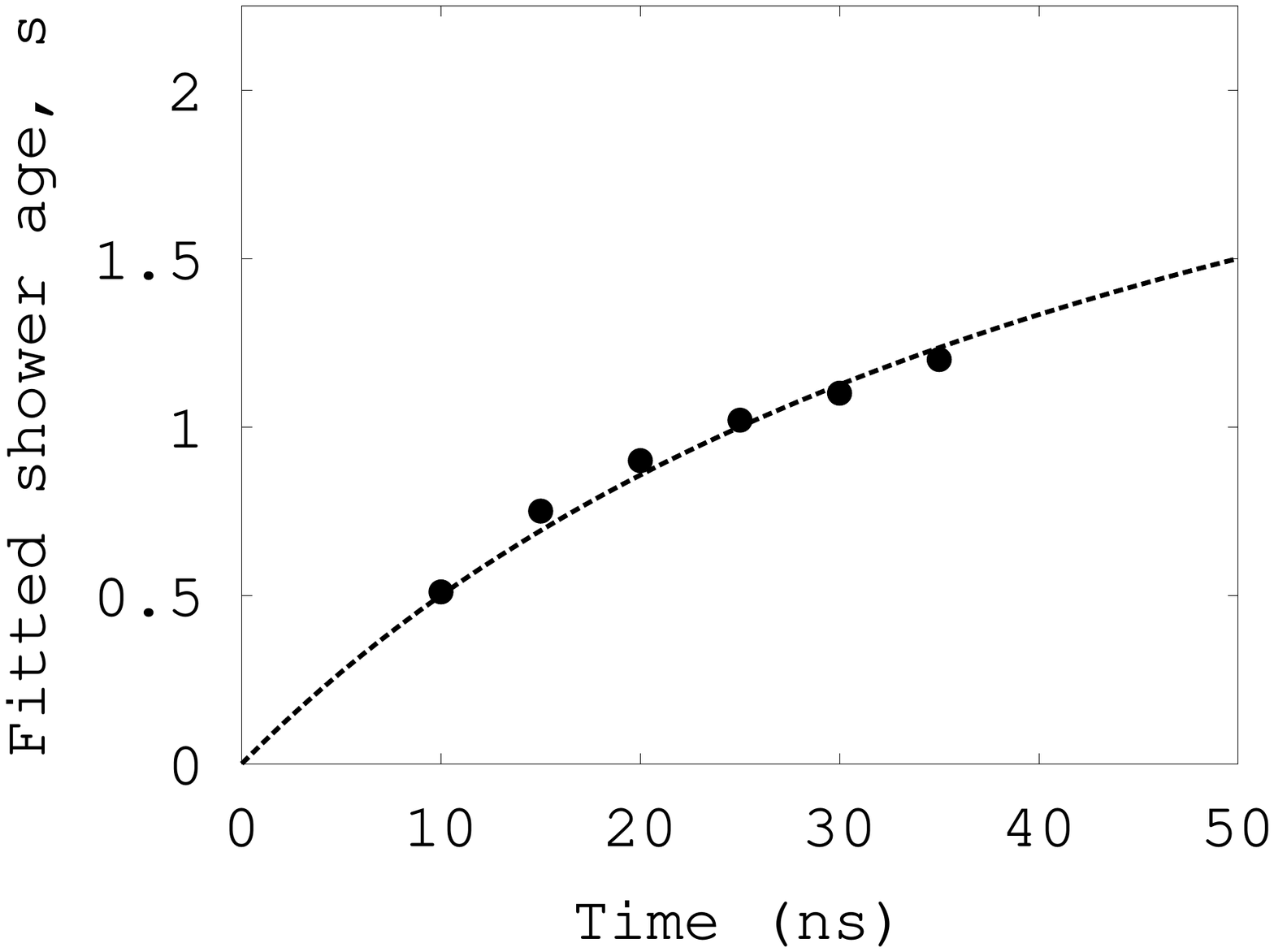}
\caption{}
\end{subfigure}
\caption{\label{fig:formG2} (a) The charged ICD density, at 10 ns and 25 ns after the first interaction.  The solid lines are fits of Equation \ref{eq:formG3} to the points with the shower age, $s$, as free parameter. (b) The fitted shower age, $s$, versus time since first interaction.  The dashed line is the theoretical expectation.}
\end{figure}

The ICD per unit area, vs. Moli\`{e}re radius, is shown in Fig. \ref{fig:formG2} (a), plotted along with Eq. \ref{eq:formG3}.  Figure \ref{fig:formG2} (b) shows the fitted shower age $s$ as a function of time after the first interaction.  Eq. \ref{eq:formG3} was fit to the MC data sets at each time bin, with $s$ as a free parameter.  The results match the definition of $s$, from which the gray dashed line in Fig. \ref{fig:formG2} is derived.

\begin{figure}
\centering
\begin{subfigure}{0.3\textwidth}
\centering
\includegraphics[width=\textwidth,trim=0cm 1cm 0cm 2cm,clip=true]{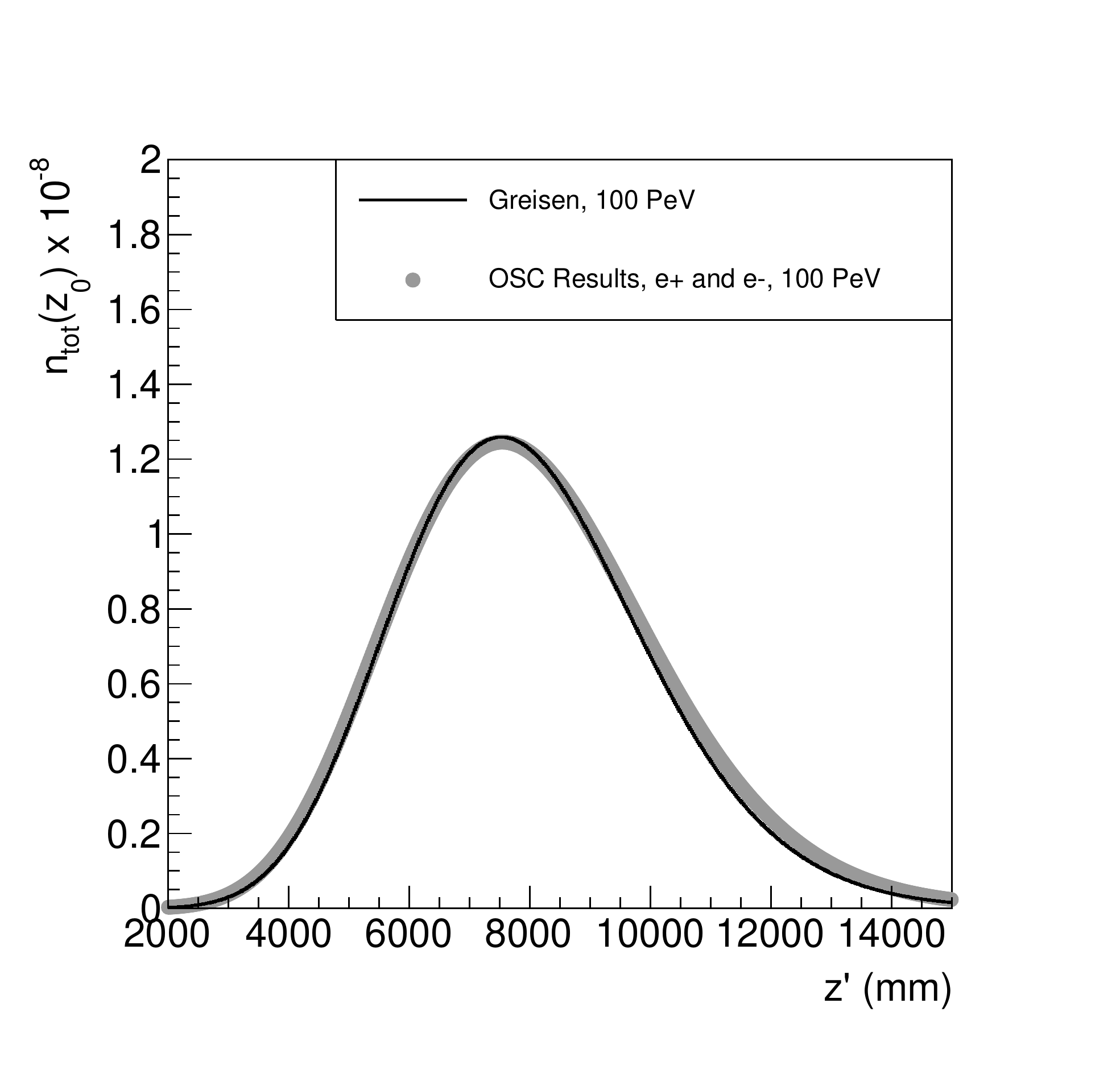}
\caption{}
\end{subfigure}%
\begin{subfigure}{0.32\textwidth}
\centering
\includegraphics[width=\textwidth,trim=0cm 1cm 0cm 2cm,clip=true]{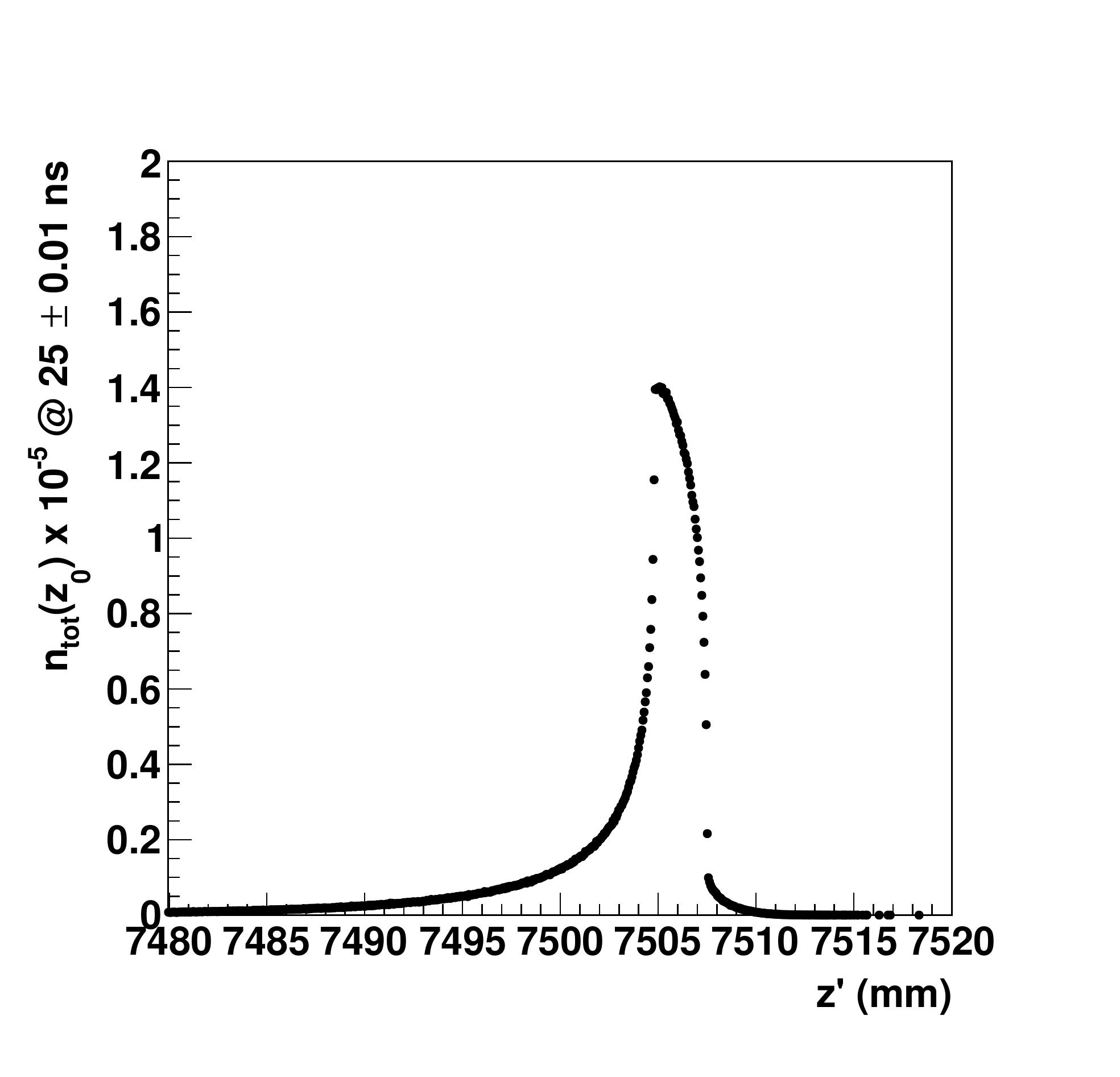}
\caption{}
\end{subfigure}%
\begin{subfigure}{0.35\textwidth}
\centering
\includegraphics[width=\textwidth,trim=1cm 2cm 1cm 3cm,clip=true]{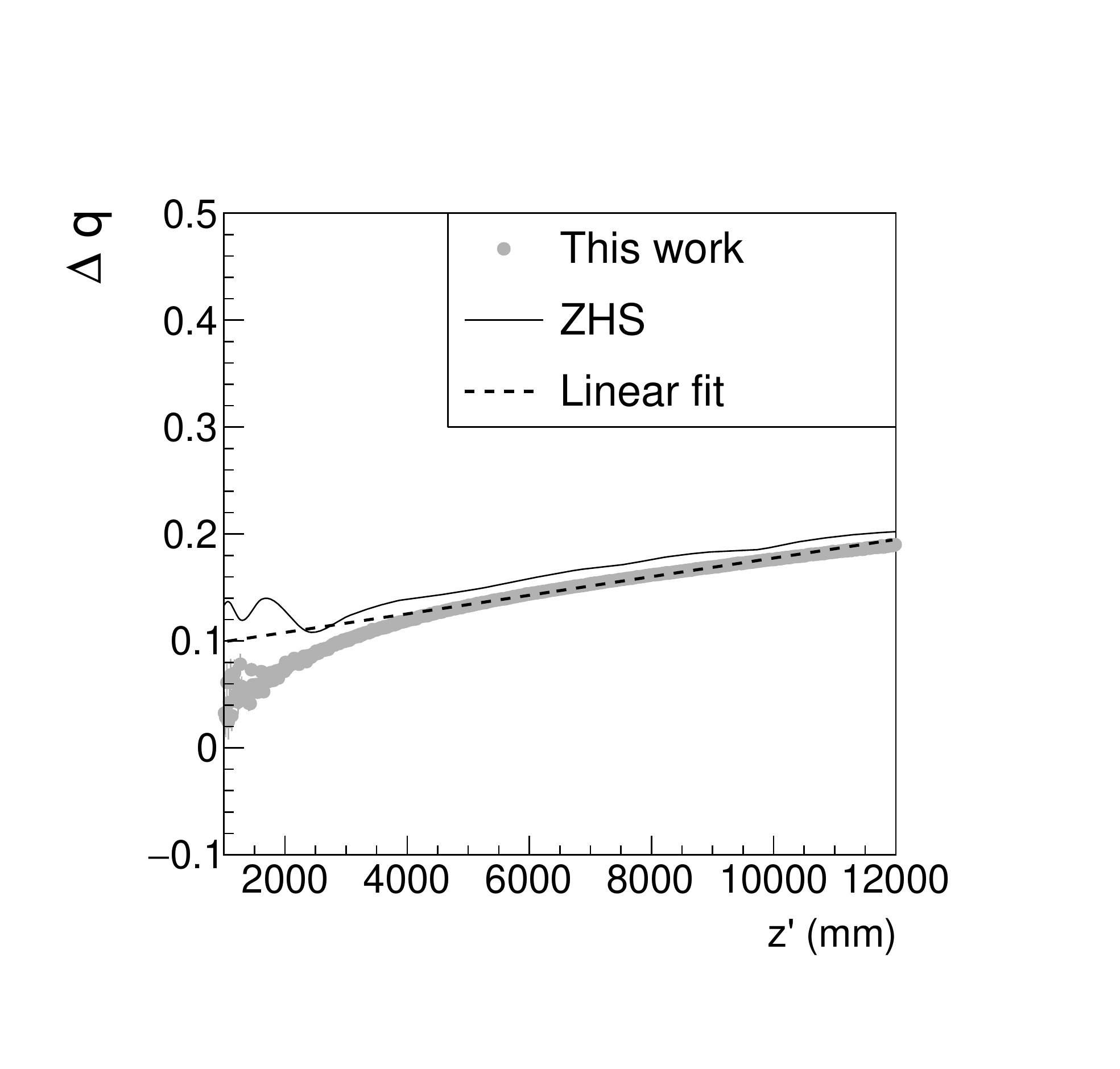}
\caption{}
\end{subfigure}
\caption{\label{fig:formG3} (a) $n_{\rm tot}$ versus $z'$, for a 100 PeV cascade.  (b) The ICD at $25\pm0.01$ ns after the first interaction. (c) The fractional negative charge excess of a 100 PeV shower, with a 5 MeV MC threshold from Geant4.  The solid line is the ZHS result with a 5 MeV MC threshold, and the dashed line is a linear fit to the OSC results.}
\end{figure}

Figure \ref{fig:formG3} (a) matches Eq. \ref{eq:formG1} to MC data, neglecting photons, with a 1 MeV MC threshold.  The Gaussian form is evident \cite{rossi1952high}, justifying the RB saddle-point expansion.  The ICD as a function of $z'$ is shown in Fig. \ref{fig:formG3}b.  The width of $f(\mathbf{x}')$ versus $z'$ is proportional to the width of the time-window (10 ps), justifying its approximation as a $\delta$-function in $f(\mathbf{x}')$.

The parameter $n_{\rm max}$ in RB is the number of excess \textit{negative} charges.  The fractional excess charge is $\Delta q = (N_{\rm e^{-}} - N_{\rm e^{+}})/(N_{\rm e^{-}} + N_{\rm e^{+}})$, so $n_{\rm max} = N \Delta q$.  The MC shows that $\Delta q$ to is linear with depth.  The y-intercept is sensitive to the MC threshold, but the slope is not.  The associated code includes the linear dependence of $\Delta q$ on depth by sampling the linear fit at $z_{\rm max}$.  Figure \ref{fig:formG3} (c) shows $\Delta q$ and that of ZHS.

\clearpage

\section*{References}

\bibliography{thebibliography}

\end{document}